\documentclass[3p]{elsarticle}

\usepackage{amssymb}
\usepackage{amsmath}
\usepackage{lineno}
\usepackage{graphicx}
\usepackage{subcaption}
\usepackage{threeparttable}
\usepackage{caption}
\journal{Mechanical Systems and Signal Processing}
\begin{document}

\begin{frontmatter}

%% Title, authors and addresses

%% use the tnoteref command within \title for footnotes;
%% use the tnotetext command for theassociated footnote;
%% use the fnref command within \author or \address for footnotes;
%% use the fntext command for theassociated footnote;
%% use the corref command within \author for corresponding author footnotes;
%% use the cortext command for theassociated footnote;
%% use the ead command for the email address,
%% and the form \ead[url] for the home page:
%% \title{Title\tnoteref{label1}}
%% \tnotetext[label1]{}
%% \author{Name\corref{cor1}\fnref{label2}}
%% \ead{email address}
%% \ead[url]{home page}
%% \fntext[label2]{}
%% \cortext[cor1]{}
%% \address{Address\fnref{label3}}
%% \fntext[label3]{}

\title{Low-pass-filter-based shock response spectrum and the evaluation method of transmissibility between equipment and sensitive components interfaces}

%% use optional labels to link authors explicitly to addresses:
%% \author[label1,label2]{}
%% \address[label1]{}
%% \address[label2]{}

\author{Yinzhong Yan}
\author{Q.M. Li\corref{cor1}}
\cortext[cor1]{Corresponding author.}
\ead{qingming.li@manchester.ac.uk}

\address{School of Mechanical, Aerospace and Civil Engineering, The University of Manchester, Manchester M13 9PL, United Kingdom}

\begin{abstract}
%% Text of abstract
According to the features of the sources of pyroshock and ballistic shock, this study considers the pyroshock and ballistic shock generated by their respective impulsive sources as damped harmonic waves with different frequencies.
According to the linear superposition assumption of damped harmonic waves in a linear elastic structure, a shock analysis method based on low-pass-filtered shock signals and their corresponding shock response spectrum (SRS), termed as low-pass-filter-based shock response spectrum (LPSRS), is proposed.
LPSRS contains rich information of the frequency distribution of the shock excitation signal.
A method to calculate shock transmissibility is proposed based on LPSRS and basic modal information of the equipment structure.
LPSRS and SRS curves can be predicted at any given position of the equipment structure.
The prediction method is validated by finite element method (FEM) simulation.

\end{abstract}

\begin{keyword}
%% keywords here, in the form: keyword \sep keyword

%% PACS codes here, in the form: \PACS code \sep code

%% MSC codes here, in the form: \MSC code \sep code
%% or \MSC[2008] code \sep code (2000 is the default)

Shock Response Spectrum \sep Low-pass Filter \sep Shock Transmissibility\sep Pyroshock\sep Ballistic Shock  

\end{keyword}

\end{frontmatter}

%% \linenumbers

%% main text
\section{Introduction}

%% The Appendices part is started with the command \appendix;
%% appendix sections are then done as normal sections
%% \appendix

%% \section{}
%% \label{}

%% If you have bibdatabase file and want bibtex to generate the
%% bibitems, please use
%%
%%  \bibliographystyle{elsarticle-num} 
%%  \bibliography{<your bibdatabase>}

%% else use the following coding to input the bibitems directly in the
%% TeX file.

A shock is a transient mechanical loading with high frequency and high amplitude.
The shocks in military and aerospace engineering sectors are called ballistic shock and pyroshock, respectively\cite{810G,ECSS2015}.
These two types of shock normally do not cause damage to main structures of armoured vehicles and spaceship.
However they may result in major functional failure of electronic and optical components, which may subsequently result in the total or partial loss of a mission.
Most shock designs and test methods are usually provided based on shock response spectrum (SRS), since a shock measurement in time domain is inconvenient for engineering applications.
SRS is generally described by the maximum absolute transient response of a single degree of freedom (SDOF) oscillator under given base excitation.
It allows to characterize the shock effect on the response of a series of SDOF oscillators in frequency domain in order to estimate its severity.
Different shocks can be compared in terms of their SRS curves, and an equivalence can be established between a real field shock and a simple shock produced in laboratory environment in terms of their SRS curves\cite{ECSS2015}.

\begin{figure*}
	\centering
	\includegraphics[width=0.6\textwidth]{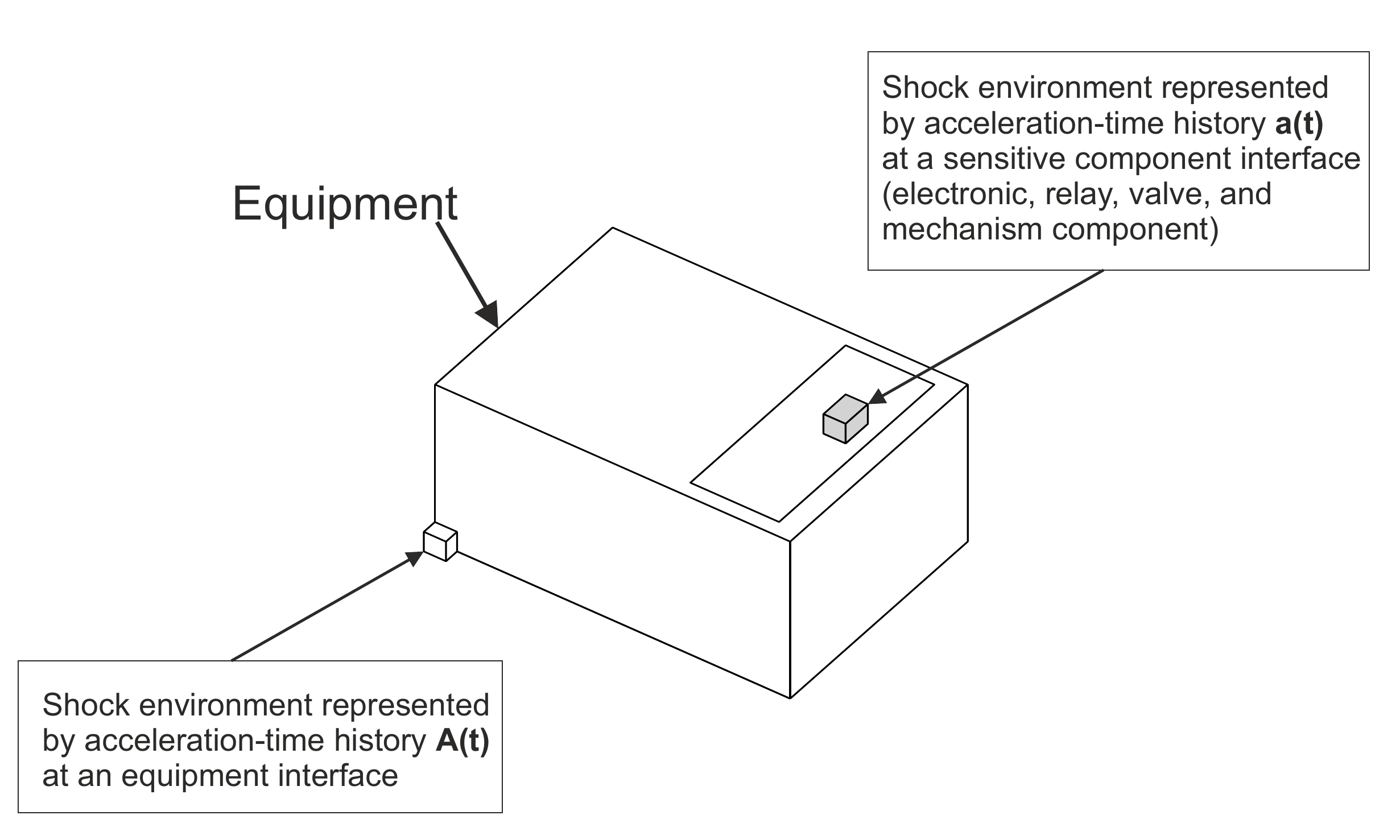}
	\caption{Schematic diagram of equipment and component interfaces}
	\label{equipment_shock_transmissibility}
\end{figure*}

One of the issues associated with a shock test is to evaluate the shock environment (usually described by SRS) at the interface of a sensitive component, which is directly related to the safety of the sensitive component in the shock environment.
A diagram describing an equipment interface and a component interface is shown in Fig.\ref{equipment_shock_transmissibility}.
Normally, measurements are only collected from equipment interface rather than component interface due to the inconvenience and uncertainty of measurement at component interface.
Thus, it is difficult to specify the exact shock environment at component interface in terms of its acceleration-time history when SRS at equipment interface is known.
Even for two different shocks with same SRS curve at equipment interface, their effects on the responses of a given component fixed to the same component interface to the equipment may differ from each other.
Especially for small units (e.g. the order of the transition frequency magnitude
\footnote{A structural response is considered as the superposition of a series of response modes.
Transition frequency is a specific frequency of the structure, beyond which higher modes can be neglected in terms of its responses\cite{Bodin2002}.}
of a printed circuit board is about 6000 Hz) under far-field shock excitation (most frequency content is below 10kHz), shock may be amplified at component interface due to resonance phenomena.
Therefore, in standards, e.g., Mechanical Shock Design and Verification Handbook, sensitive components are required to be tested with whole equipment, rather than be tested directly as stand-alone components.

With SRS curve at equipment interface and an appropriate method, e.g. square root of the sum of squares (SRSS), the upper bound response of the equipment at the component interface (which is also the upper bound excitation at component interface) can be estimated.
However, only with the upper bound excitation at component interface, the response of the component cannot be calculated.
European Cooperation for Space Standardisation (ECSS) has used the shock transmissibility between equipment and component interfaces obtained from sine sweep tests (up to 2000 Hz) to determine the shock environment at component interface from the shock environment at equipment interface.
Then a 6 dB corridor between SRS curves at component and equipment interfaces from 2000 Hz to a transition frequency is assumed.
According to the ECSS handbook\cite{ECSS2015}, this evaluation method relies mainly on rules-of-thumb, which cannot be considered as a reliable method.
The use of a smaller corridor may lead to material failure, while the use of a larger corridor may result in increases of weight, design period and material cost.
Therefore, a new and reliable shock transmissibility evaluation method based on shock propagation mechanism is necessary.

This study considers pyroshock and ballistic shock as the superposition of damped harmonic waves based on the feature of their shock generation mechanisms.
In this case, it is proposed that the superposition of SRS amplitudes in different frequency bands, i.e., linear superposition of amplitudes, can be satisfied.
With a low-pass-filter-based SRS (LPSRS) proposed in this paper, the shock environment at component interface can be predicted with some basic modal information of the equipment structure.
LPSRS method for shock transmissibility also provides a theoretical support for laboratory shock test, with which it is possible to use simple shocks to test stand-alone sensitive components, rather than testing the sensitive components on whole equipment.
The predicted result is validated by finite element method (FEM) simulation results.

\section{LPSRS and Transmissibility Evaluation Method}
\subsection{Linear Superposition Assumption}
\label{Section_Assumption}

\begin{figure}
	\centering
	\includegraphics[width=0.5\linewidth]{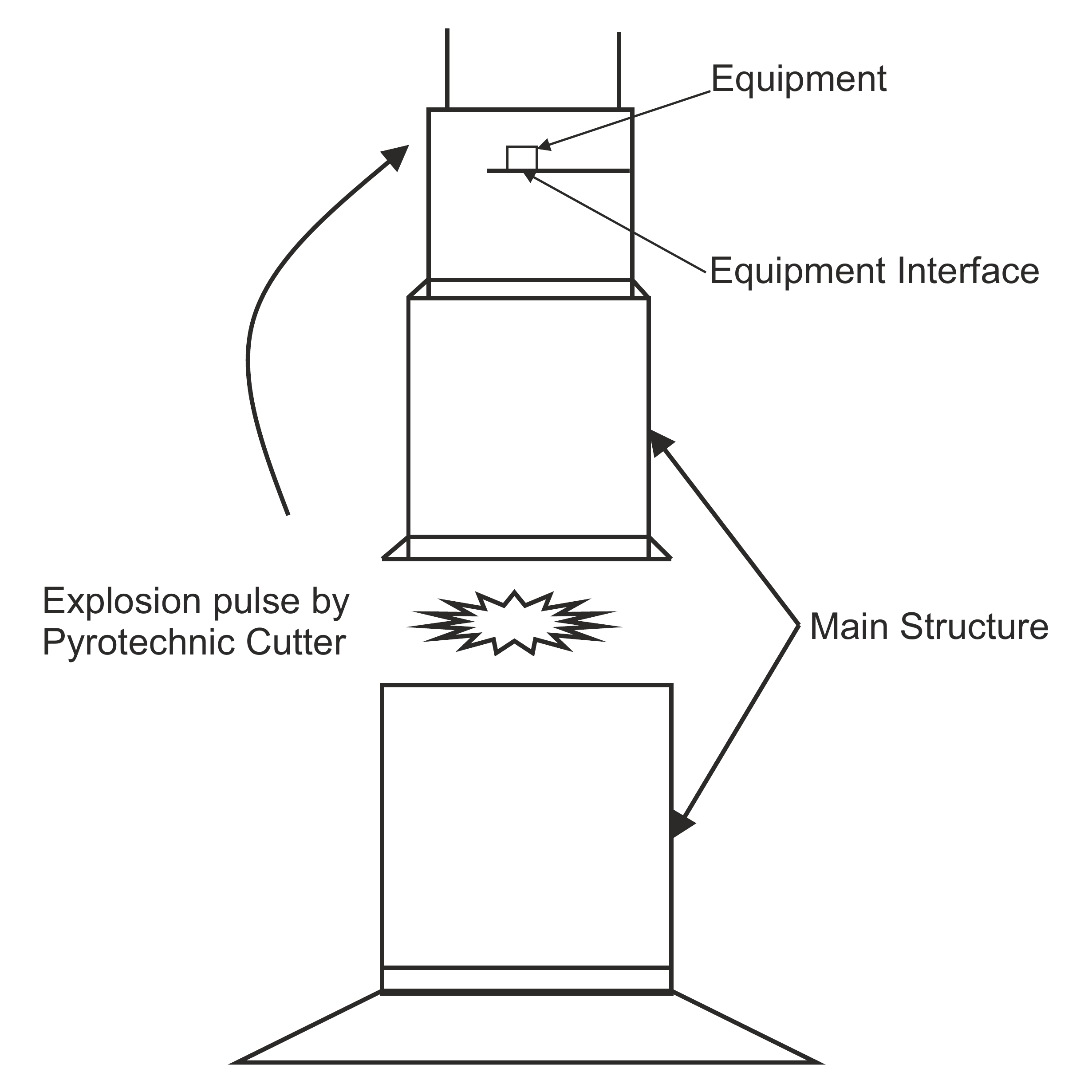}
	\caption{Shock transmission from a shock source to an equipment interface}
	\label{transmission_main_structure}
\end{figure}

Before analysing shock transmissibility from equipment interface to component interface, it is necessary to understand the shock generation process and the basic features of pyroshock and ballistic shock.
Pyroshock is a specific shock which is caused by the detonation of pyrotechnic devices, while ballistic shock is caused by the impact between a kinetic projectile and structure with possible involvement of blast effects.
Fig.\ref{transmission_main_structure} is a schematic diagram to illustrate shock generation and its transmission from shock source to equipment interface for a rocket.

These two types of shock sources at loading location where explosion and/or impact occur can be represented by a simple pulse load (e.g. sinusoidal or rectangular force-time history) with some secondary high frequency oscillations that decays rapidly during propagation.
According to the example given in \ref{section_appendix}, the response of a SDOF oscillator under a simple pulse load can be considered as a damped harmonic wave with maximum acceleration at the starting point in the forced response phase.
Therefore the pyroshock and ballistic shock environment at an equipment interface is mainly the superposition of damped vibration modes of the main structure.

However, the shock signal at equipment interface is the superposition of a series of damped harmonic waves with a wide range of frequencies determined by the modal responses of the main structure.
The complexity mainly comes from high frequency density of higher modes, which can be easily excited by pyroshock and ballistic shock.
The purpose of linear superposition assumption introduced below is to separate a shock into limited terms in a complete range of frequency band windows, which is based on the requirement that the time $\tau_i$ at the peak of each band-pass-filtered term happens around the same time $T$.

\begin{table*}
	\centering
	\caption{Definition of symbols}
	\label{symbol_definition}
	\begin{tabular}{p{0.15\textwidth} p{0.75\textwidth}}
		\hline
		Symbols & Definitions \\
		\hline
		$A(t),A$ & Acceleration-time history and its maximum-absolute amplitude at equipment interface\\
		$A_i(t),\,A_i$ & Acceleration-time history and its maximum-absolute amplitude of band-pass-filtered acceleration signal with band frequency range of $(f_{i-\frac{1}{2}},\,f_{i+\frac{1}{2}})$ at equipment interface \\
		$A^i(t),\,A^i$ & Acceleration-time history and its maximum-absolute amplitude of low-pass-filtered acceleration signal with cut-off frequency $f_{i+\frac{1}{2}}$ at equipment interface \\
		$A_f$ & Amplitude of absolute acceleration SRS of $A(t)$ at frequency $f$ at equipment interface\\
		$A_{if}$ & Amplitude of absolute acceleration SRS of $A_i(t)$ at frequency $f$ at equipment interface\\
		$A_n$ & $A_n=A_f$ when $f=\frac{\omega_n}{2\pi}$ for the nth vibration mode of the equipment\\
		$A_{in}$ & $A_{in}=A_{if}$ when $f=\frac{\omega_n}{2\pi}$ for the nth vibration mode of the equipment\\
		$a(t),a$ & Acceleration-time history and its maximum-absolute amplitude at component interface\\
		$a_i(t),\,a_i$ & Acceleration-time history and its maximum-absolute amplitude of band-pass-filtered acceleration signal with band frequency range of $(f_{i-\frac{1}{2}},\,f_{i+\frac{1}{2}})$ at component interface\\
		$a_f$ & Amplitude of absolute acceleration SRS of $a(t)$ at frequency $f$ at component interface\\
		$a_{if}$ & Amplitude of absolute acceleration SRS of $a_i(t)$ at frequency $f$ at component interface\\
		\hline
	\end{tabular}
\end{table*}

For a shock excitation of an acceleration-time history $A(t)$, a series of band-pass-filtered signals can be generated, i.e., $A_{i}(t)$ is a band-pass-filtered signal of $A(t)$ with $i$th band frequency range of $(f_{i-\frac{1}{2}},f_{i+\frac{1}{2}})$.
With an ideal band-pass-filter, filtered signals have following relation
\begin{equation}
\sum_{i=1}^N A_i(t)=A(t)
\end{equation}
where $FFT(A(t))=0$ when the frequency $f\leq f_{\frac{1}{2}}$ and $f \geq f_{N+\frac{1}{2}}$.

Linear superposition assumption is stated as
\begin{equation}
	\label{a_sum}
	\sum_{i=1}^N A_{i}=A
\end{equation}
where $A_i=max|A_i(t)|$, and $A=max|A(t)|$ are the maximum-absolute amplitude (amplitude is used to represent absolute amplitude in the rest of paper) of the acceleration signal filtered by the $i$th band-pass filter and the maximum-amplitude of the original pyroshock or ballistic shock, respectively.

Linear superposition assumption is valid when $\tau_i=T$, i.e.
\begin{equation*}\label{linear_superpostion_assumption}
\sum_{i=1}^N max|A_i(t)|=\sum_{i=1}^N |A_i(\tau_i)|=\sum_{i=1}^N |A_i(T)|=max|\sum_{i=1}^N A_i (t)|=max|A(t)|
\end{equation*}
It should be noted that the maximum-absolute function is used to be consistent with the usually-used absolute acceleration SRS.
Alternatively, positive acceleration SRS or negative acceleration SRS can also be used, which are approximately equal to absolute acceleration SRS for ballistic shock and pyroshock\cite{Irvine2002}.
The maximum-absolute function is usually a non-linear function unless the peak of every term happens around the same time.
Therefore, Eq.(\ref{a_sum}) is not always true because the peak time of each band-filtered term does not always synchronise.
However by selecting appropriate pass band, $A_i(t)$ terms are possibly to reach their peaks around the same time.
For pyroshock/ballistic shock environment, pass band at every octave frequency is suggested, and this selecting criterion can be further optimised based on the particular shock signal concerned. 

This linear superposition assumption can be extended to the shock responses of a SDOF model to $A(t)$, i.e.
\begin{equation}
	\label{ani_sum}
	\sum_{i=1}^N A_{if}=A_f
\end{equation}
and to the response of the component interface to $A(t)$, i.e.
\begin{equation}
	\label{Ai_sum}
	\sum_{i=1}^N a_i=a
\end{equation}
The definitions of the symbols in Eq.(\ref{ani_sum}) are summarized in Table \ref{symbol_definition}.

\subsection{Low-Pass-Filter-Based Shock Response Spectrum}
For any SDOF model subjected to a damped harmonic base oscillation with amplitude $A_0$, damping ratio $\bar{\xi}$ and frequency $\bar{\omega}$, its dynamic equation of relative motion is
\begin{equation}
\label{damped_motion}
\ddot{v}(t)+2\xi\omega\dot{v}(t)+\omega^{2}v(t)=A_{0} \, e^{-\bar{\xi} \bar{\omega} t} \, sin\, \bar{\omega}t 
\end{equation}
with following relative displacement and absolute acceleration solutions
\begin{equation}
v(t)=\frac{A_{0}}{\omega_D}\int_{0}^{t} \, e^{-\bar{\xi} \bar{\omega} \tau} \, \sin( \bar{\omega}\tau)\,e^{-\xi\omega(t-\tau)}\,\sin(\omega_D(t-\tau))\,d\tau
\end{equation}
\begin{equation}\label{duhamel_solution}
\ddot{z}(t)=\dfrac{A_0}{\omega_D}\int_{0}^{t} e^{-\bar{\xi} \bar{\omega} \tau} \, \sin( \bar{\omega}\tau)e^{-\xi\omega (t-\tau)}[(\omega_D^2-\xi^2\omega^2)\sin(\omega_D (t-\tau))+2\xi \omega \omega_D \cos(\omega_D t-\tau)]d\tau
\end{equation}
where $\omega$ is oscillator's circular natural frequency; $\xi$ is its damping ratio and $\omega_D=\omega \sqrt{1-\xi^2}$ is its damped natural frequency.

\begin{figure}
	\centering
	\includegraphics[width=0.7\textwidth]{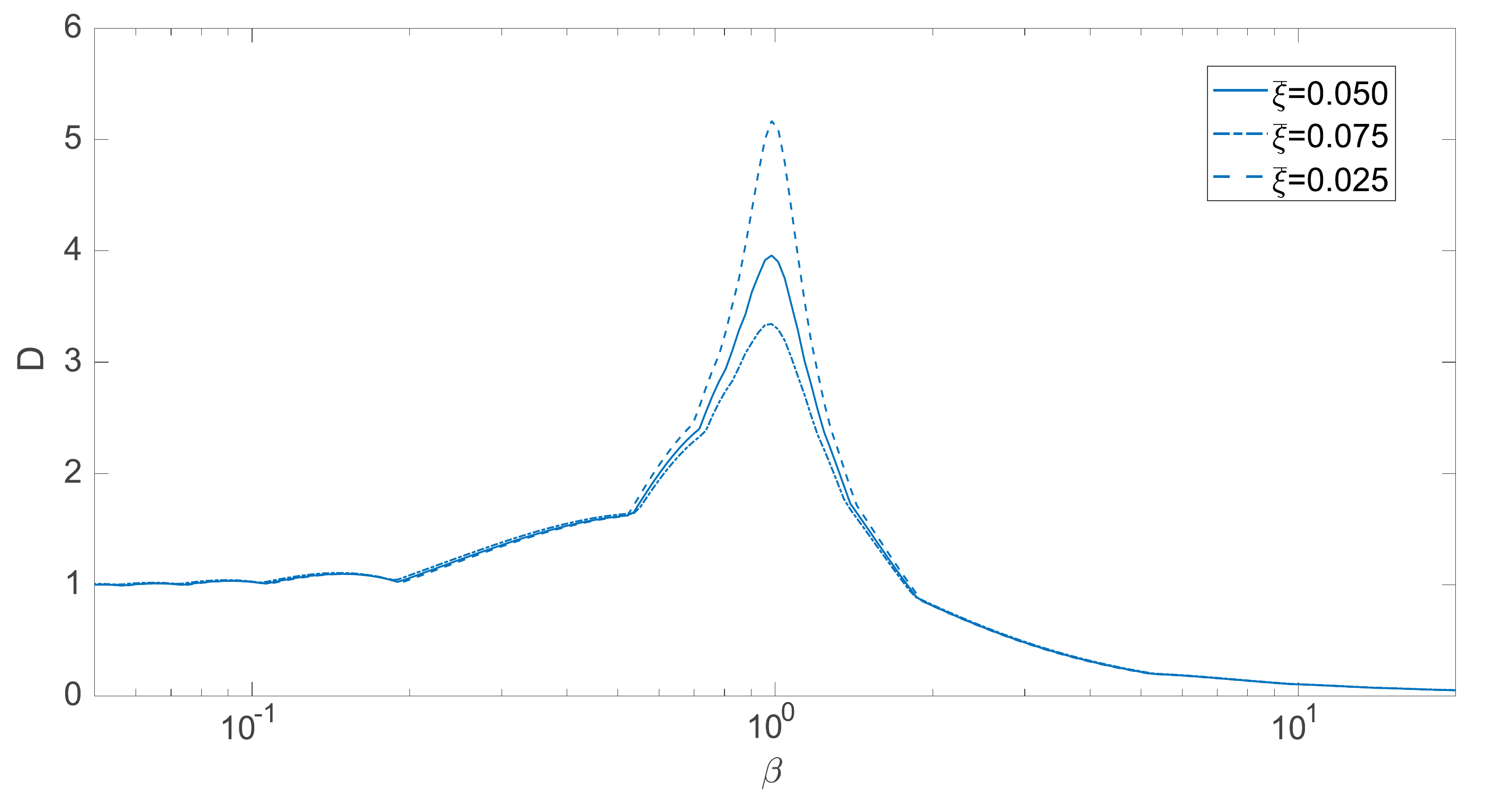}
	\caption{Dynamic amplification ratio of the acceleration of a SDOF model when $\xi=0.05$ under damped harmonic excitation with $\bar{\xi}$=0.025, 0.050, 0.075.}
	\label{damped_dynamic_amplification}
\end{figure}

Fig.\ref{damped_dynamic_amplification} shows the dynamic amplification ratio $D$, i.e., the maximum response acceleration normalized by its maximum excitation acceleration, under damped harmonic base excitation with different damping ratios and a range of frequency ratio $\beta$ (defined as $\frac{\bar{\omega}}{\omega}$) for $\xi=0.05$.
From this plot, it can be seen that the acceleration response of a SDOF oscillator is approximately equal to the amplitude of excitation signal when the excitation frequency is much lower than the natural frequency of SDOF ($\beta\rightarrow0$)\footnote{When the excitation frequency is much higher than the natural frequency of SDOF ($\beta\gg1$), $D\rightarrow0$.}.
Based on this equality, the LPSRS method is developed, which is illustrated below.

\begin{figure}
		\centering
	\begin{subfigure}[b]{\textwidth}
		\includegraphics[width=0.68\textwidth]{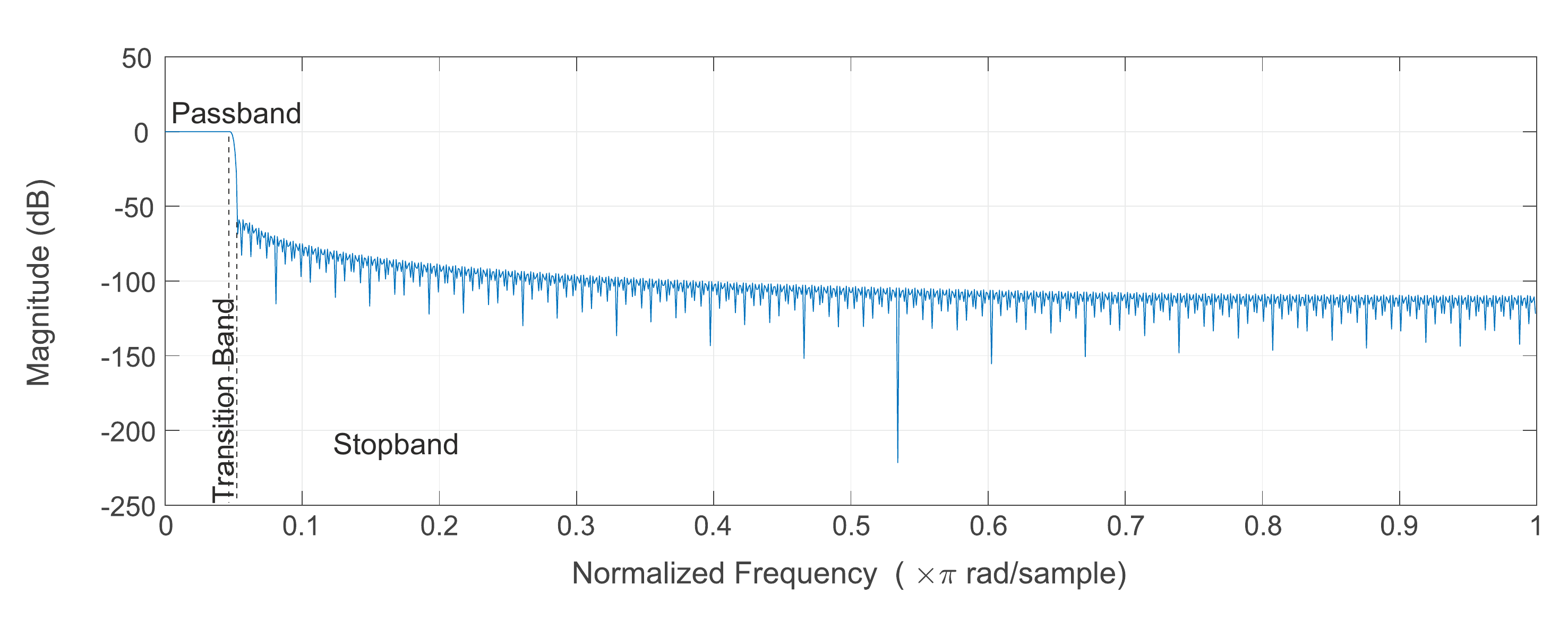}
		\centering
		\caption{}
		\label{frequency_response}
	\end{subfigure}
	\begin{subfigure}[b]{\textwidth}
		\includegraphics[width=0.68\textwidth]{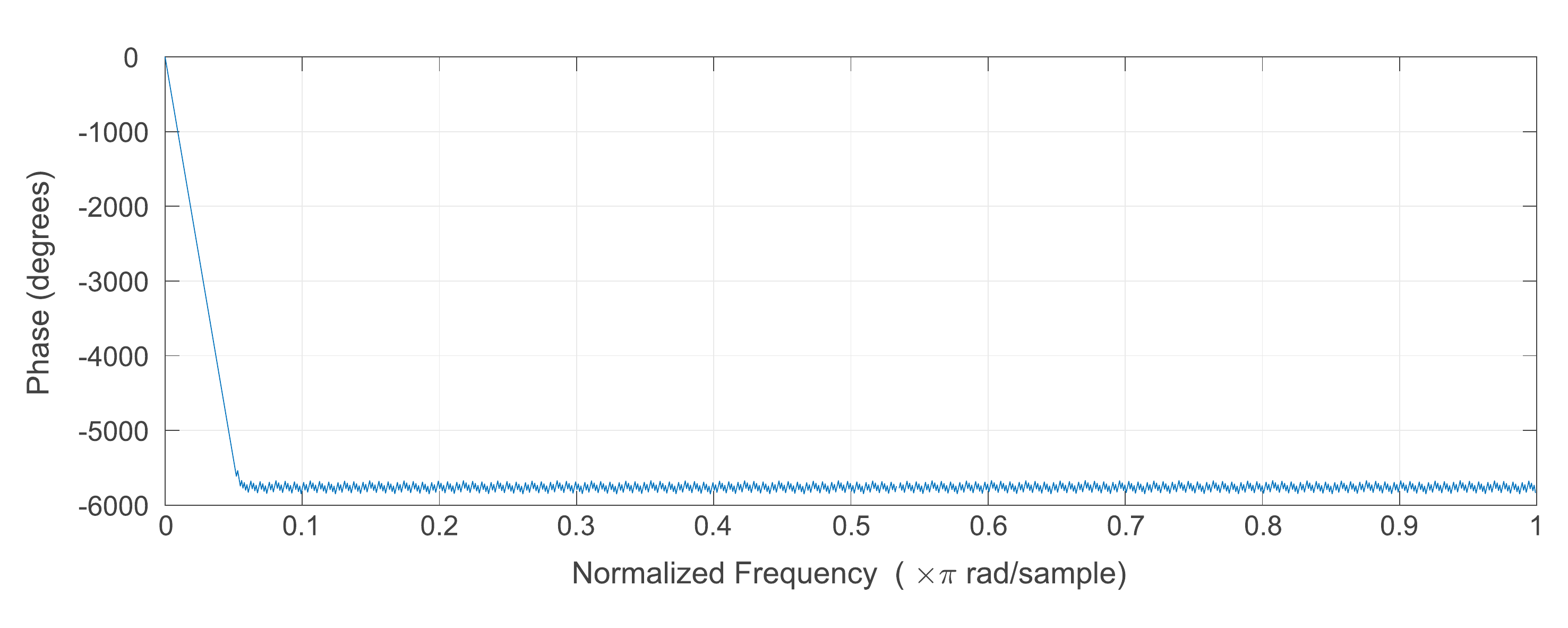}
		\centering
		\caption{}
		\label{phase_response}	
	\end{subfigure}
	\caption{(a) Magnitude response, and (b) phase responses, of the FIR filter (Bode plot)}
	\label{bode_plot}
	\end{figure}

Low-pass filters are applied to the base excitation signal at equipment interface to identify the frequency distribution.
In this analysis, low-pass Finite Impulse Response (FIR) filters based on Hamming window are used.
Low-pass cut-off frequencies are selected in an equidistant log scale, e.g., every octave in this paper.
They may be chosen in other forms of cut-off frequencies according to the particular problem, which, however, is outside the research focus of this study.
With this FIR filter, the transition from passband to stopband is rapid while passband ripple remaining sufficiently small.
The response curve (Bode Plot) of this Hamming-window-based FIR digital filter is shown in Fig.\ref{bode_plot}.

\begin{figure}
	\centering
	\includegraphics[width=0.8\textwidth]{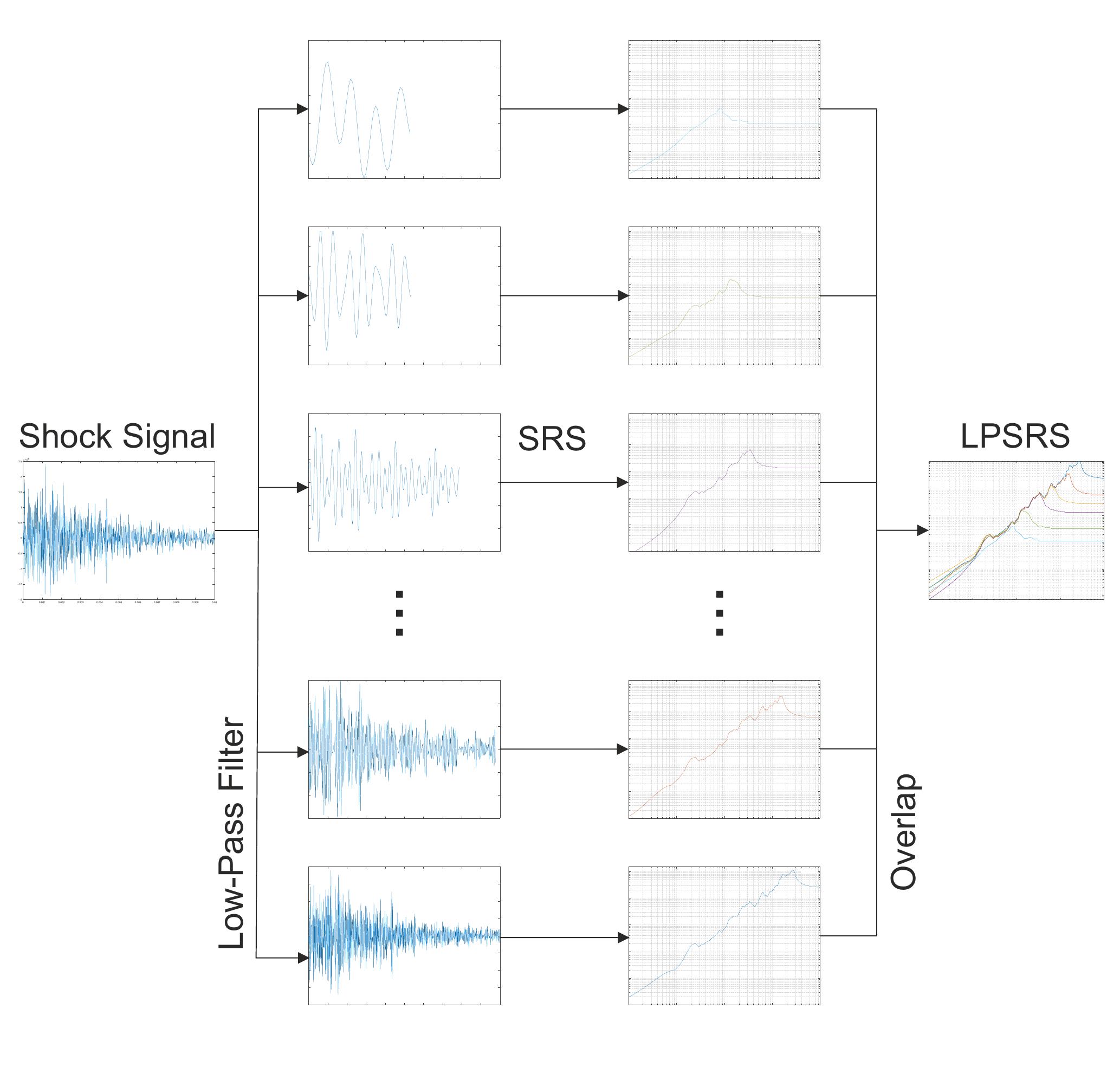}
	\caption{Flowchart of the proposed Low-Pass Filter Based SRS (LPSRS)}
	\label{LPSRS_process}
\end{figure}

Fig.\ref{LPSRS_process} shows the flowchart of the proposed LPSRS method.
The method begins with applying low-pass filter to the base excitation signal at each cut-off frequency $f_{i+\frac{1}{2}}$.
Then the SRS of each filtered signal is plotted.
An important requirement for plotting LPSRS is to plot each SRS till the appearance of its frequency-independent platform.
These platforms are also called temporal peak in \cite{Hwang2016}.
According to Fig.\ref{damped_dynamic_amplification}, if the natural frequency of a SDOF oscillator is much higher than the excitation frequency $f_{i+\frac{1}{2}}$, the amplitude of the SDOF response is equal to the maximum amplitude of the excitation $A^i$, and will stay the same no matter how SDOF natural frequency increases, which underpins the existence of the frequency-independent platform in an SRS plot.
Normally SRS platform appears when the frequency on SRS plot (i.e. the frequency of the SDOF oscillator) is five times higher than the cut-off frequency of the filter (i.e. the maximum frequency of the input excitation), as shown in Fig.\ref{damped_dynamic_amplification}.
The last step is to overlay all SRS curves obtained together into one plot, which is the LPSRS. 

From Fig.\ref{damped_dynamic_amplification}, the dynamic amplification ratio approaches to 1 with $\beta\ll1$.
Basically, dynamic amplification ratio ($D$) can be treated as 1 when $\beta$ is smaller than 0.2, so that the maximum frequency of a SRS curve needs at least to be 5 times as high as the cut-off frequency as discussed above.
According to ECSS handbook \cite{ECSS2015}, sampling rate of the signal should be at least 8 times as high as the maximum frequency for the SRS analysing frequency.
This means that sampling rates are usually more than 40 times of the cut-off frequency.
In this study, when LPSRS is constructed, the ratio between sampling rate and cut-off frequency is always kept as 40 for all cut-off frequencies.

\begin{figure}
	\centering
	\includegraphics[width=0.6\textwidth]{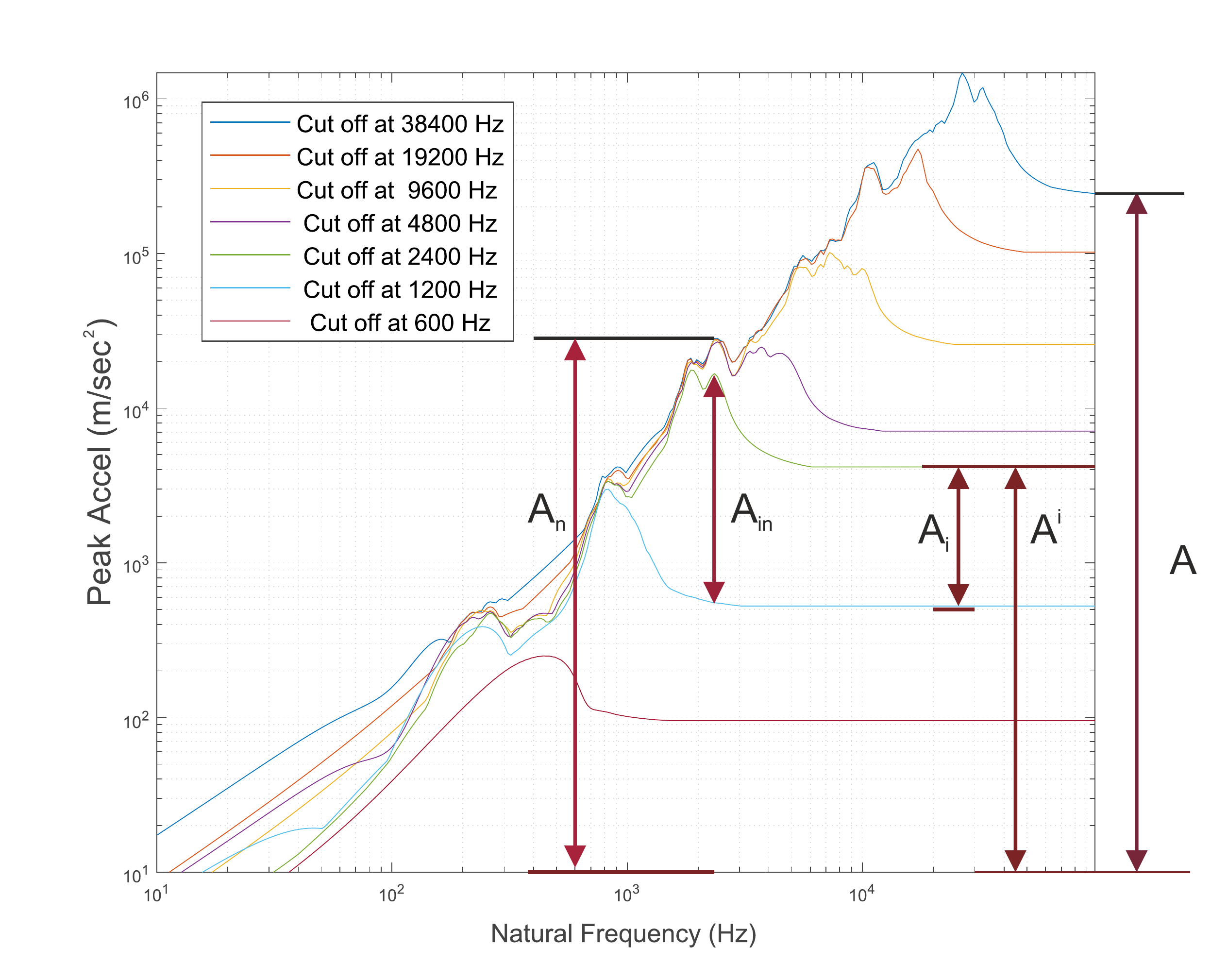}
	\caption{LPSRS example}
	\label{sample_lpsrs}
\end{figure}

With $N$ SRS curves plotted in a LPSRS, there are $N$ amplitude differences between two neighbouring SRS curves.
The band-pass-filtered signal can be calculated by
\begin{equation}\label{low_to_band_t}
A_i(t)=A^i(t)-A^{i-1}(t)
\end{equation}
where $A^i(t)$ is defined in Table \ref{symbol_definition}.

With linear superposition assumption and based on Eq.(\ref{low_to_band_t}), the maximum amplitude of band-pass-filtered signal $A_i$ can be expressed as
\begin{equation}\label{low_to_band}
A_i=A^i-A^{i-1}
\end{equation}
where $A^i$ is defined in Table \ref{symbol_definition}.

Equations (\ref{low_to_band_t}) and (\ref{low_to_band}) can be verified by obtaining $A_i(t)$ (or, $A_i$) and $A^i(t)-A^{i-1}(t)$ (or, $A^i-A^{i-1}$) independently from $A(t)$ and comparing them.

An example LPSRS is shown in Fig.\ref{sample_lpsrs}, where cut-off frequencies are selected every octave from 0.6 kHz to 38.4 kHz to produce the LPSRS curves.
Since high frequency contents of a shock have almost no influence on the response of low frequency oscillators, all SRS curves increase with frequency in the same path, which level off after passing their respective cut-off frequencies.
The analysing frequency of SRS is up to 38.4 kHz.
Hence $A$ is the maximum amplitude of the acceleration excitation signal filtered by 38.4 kHz low-pass filter.
$A_i$ is the maximum amplitude of band-pass-filtered signal $A_i(t)$, which is between $f_{i-\frac{1}{2}}$=1200 Hz to $f_{i+\frac{1}{2}}$=2400 Hz when $i=3$ as an illustration case in Fig.\ref{sample_lpsrs}.

\subsection{Transmissibility Evaluation Method}

\begin{figure}
	\centering
	\includegraphics[width=0.6\textwidth]{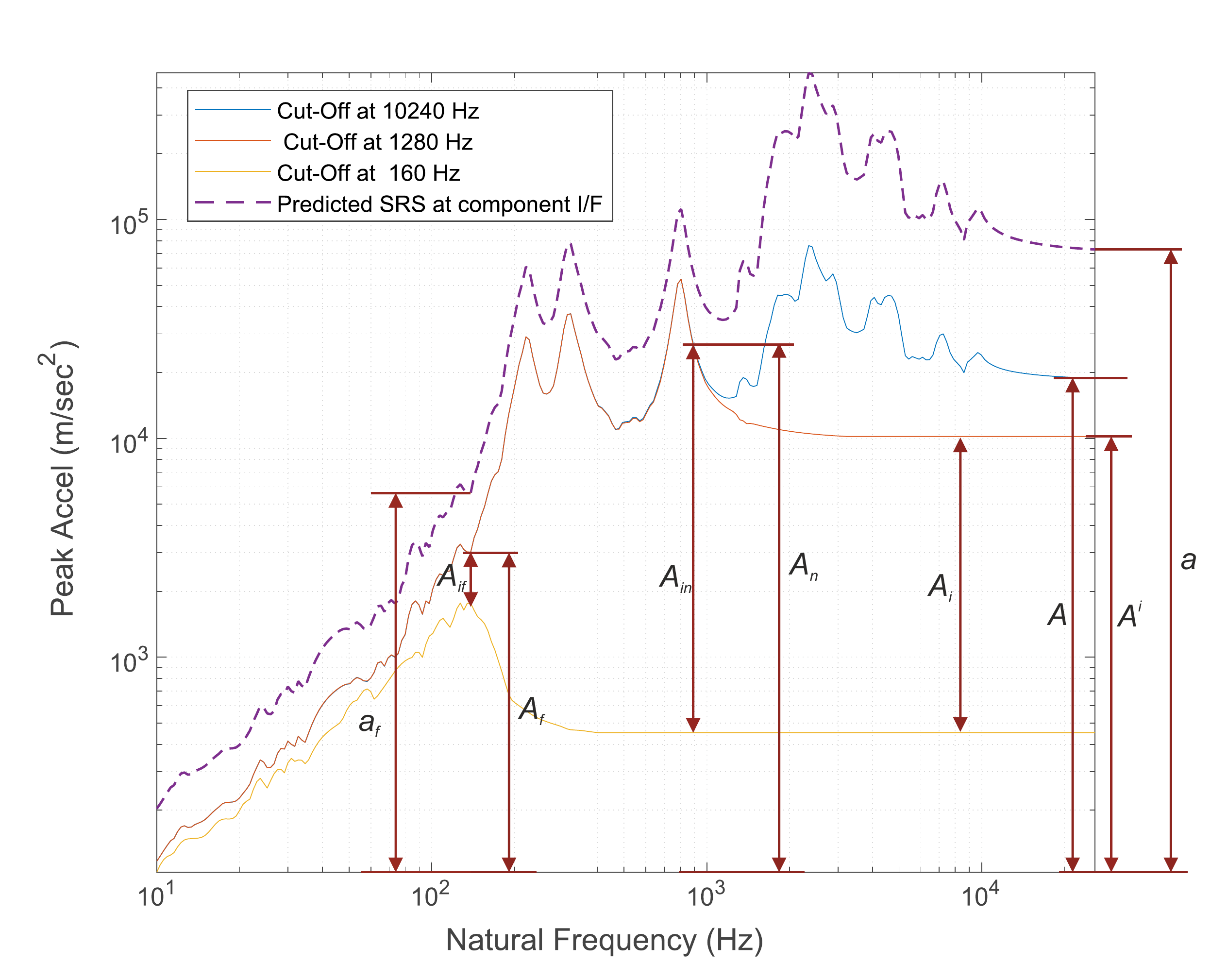}
	\caption{Schematic diagram of symbols}
	\label{evaluation_process}
\end{figure}

As shown in Fig.\ref{equipment_shock_transmissibility}, the acceleration-time histories of the shock at equipment and component interfaces are represented by $A(t)$ and $a(t)$, respectively. Fig.\ref{evaluation_process} is a diagram to illustrate symbols defined in Table \ref{symbol_definition}.

Although it is difficult to obtain the acceleration-time history at component interface, it is possible to estimate upper bound ($a^u$) of amplitude ($a$) of excitation $a(t)$ at component interface by SRS analysis methods\cite{Irvine2010,Alexander2009}, e.g. absolute sum method, i.e.

\begin{equation}
	\label{abssum}
	a^u=\sum_n |P_n \phi_n A_{n}|
\end{equation}
in which $P_n$ is modal participation factor; and $\phi_n$ is mode shape vector of mode $n$. $A_n$ is the response amplitude of the $n$th vibration mode of the equipment at equipment's natural frequency $f_n$, which can be obtained from SRS plot at equipment interface directly.
Usually, $a\leq a^u$. As a conservative estimation, taking
\begin{equation}\label{cancel_upper_bound}
a\approx a^u
\end{equation}
which has been used in Refs. \cite{Irvine2010,Alexander2009}.

Substituting Eqs.(\ref{ani_sum})\footnote{Eq.(\ref{ani_sum}) is valid when $f=\frac{\omega_n}{2\pi}$, i.e. $A_f=A_n$ as defined in Table \ref{symbol_definition}.} and (\ref{cancel_upper_bound}) into Eq.(\ref{abssum}), the shock response at component interface becomes
\begin{align}\label{full_expression}
	a & = \sum_n |P_n \phi_n| |A_{n}|\nonumber\\
	& = \sum_n |\phi_n P_n| (\sum_i |A_{in}|)\nonumber\\
	& = \sum_i \sum_n |\phi_n P_n A_{in}|
\end{align}

With linear superposition assumption, the result is independent of summation order (frequency $f_i$ and $f_n$). Based on Eqs.(\ref{Ai_sum}) and (\ref{full_expression}), $a_i$ can be obtain from modal superposition, i.e.
\begin{equation}
	a_i=\sum_n |\phi_n P_n A_{in}|
\end{equation}
From this equation, amplitude of band-pass-filtered shock signal with frequency band $(f_{i-\frac{1}{2}},f_{i+\frac{1}{2}})$ at component interface can be computed directly.

\begin{figure}
	\centering
	\includegraphics[width=0.6\linewidth]{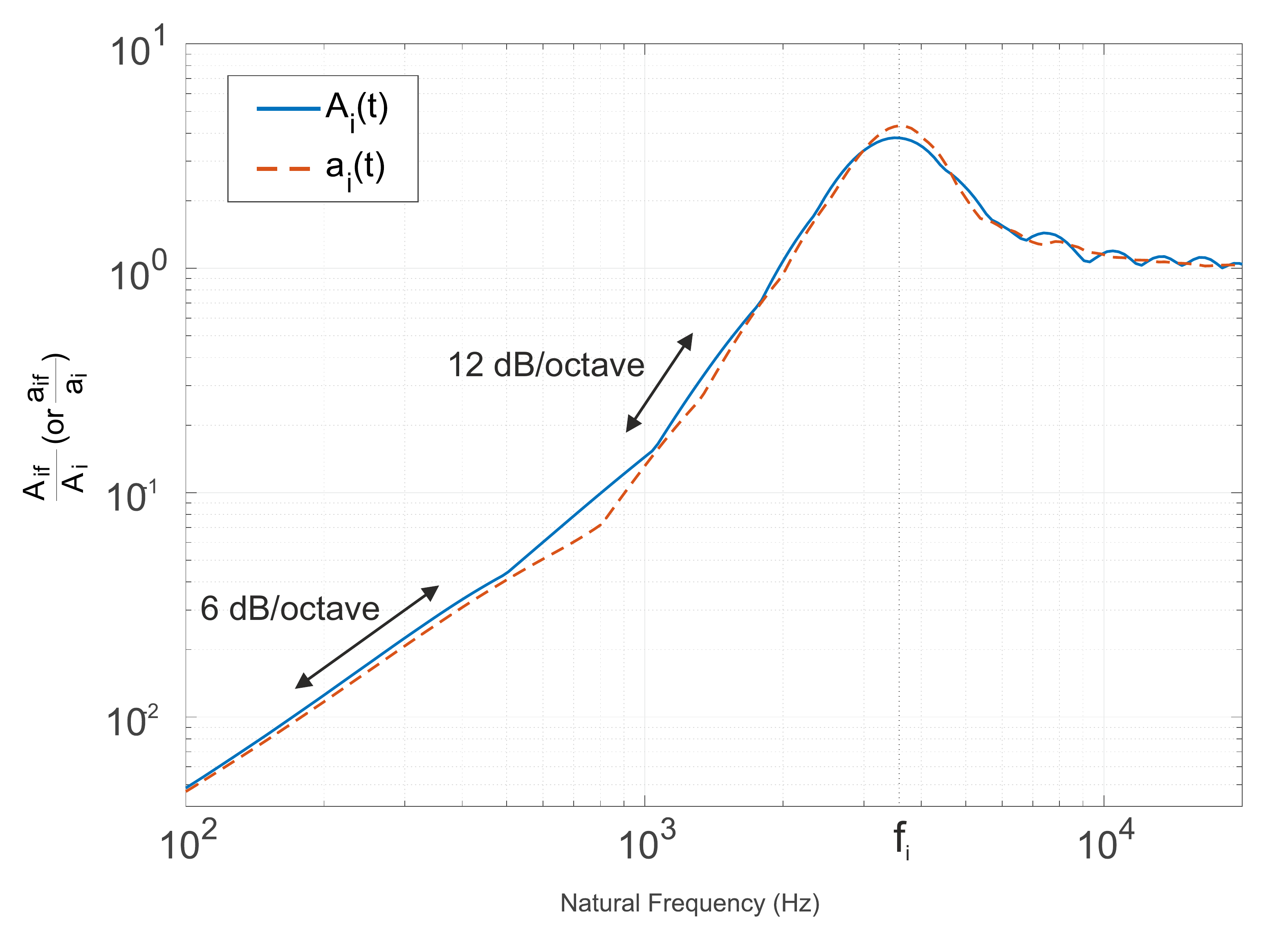}
	\caption{SRS characteristics of band-pass-filtered shock $A_i(t)$ and $a_i(t)$}
	\label{scaling_support}
\end{figure}
Based on the SRS properties given by the ECSS handbook(\cite{ECSS2015}, p.30), if the velocity change of a shock is zero, $A_{if}$(or, $a_{if}$) have an initial slope of 6 dB/octave and raising to 12 dB/octave up to the corner frequency $f_i$
\footnote{The corner frequency (also known as knee frequency) on SRS is the frequency corresponding to the peak value of $A_{if}$ (or, $a_{if}$)\cite{ECSS2015}.}.
After the SRS is normalized by $A_i$(or, $a_i$), the value of the SRS approaches to unity as the natural frequency of the SDOF exceeds corner frequency $f_i$.
The amplification factor $\frac{A_{if}}{A_i}$ and $\frac{a_{if}}{a_i}$ at corner frequency $f_i$ depend on their signal forms, which can be considered to be similar for the same shock event.
Fig.\ref{scaling_support} shows an example of these SRS characteristics, where $A_i(t)$ and $a_i(t)$ are taken from the simulation in Fig.\ref{FE_Modal} when $i=3$, to show the scaled equivalence of the normalized SRS, i.e. 
\begin{equation}\label{DA_equivalence}
\frac{a_{if}}{a_i} =\frac{A_{if}}{A_i}
\end{equation}
Thus,
\begin{align}
	a_{if} & =\frac{a_i}{A_i}A_{if}\nonumber\\
	& =\frac{\sum_n |\phi_n P_n A_{in}|}{A_i}A_{if}
	\label{frequency_information_hold}
\end{align}
Response of SDOF with different frequencies at component interface $a_f$ can be obtained from superposition of its every individual frequency content according to linear superposition assumption:
\begin{align}
	a_f & =\sum_i a_{if}\nonumber\\
	& =\sum_i \frac{\sum_n |\phi_n P_n A_{in}|}{A_i}A_{if}
	\label{LPSRS_algorithm}
\end{align}
By plotting $a_f$ and $a_{if}$ against frequency variable $f$, SRS and its LPSRS plots at component interface are obtained.
With original SRS and spectrum analysis method, only upper bound of the acceleration excitation at component interface can be derived.
While with $a_f$ and $a_{if}$ values calculated from LPSRS, much more information of shock severity at component interface can be obtained than that obtained from the upper bound acceleration excitation.
The information contained in a LPSRS graph is sufficient to estimate shock transmissibility in a structure.
Example of this predictive method is given in Section \ref{numerical_simulation_chapter}.

\section{Numerical Simulation and Result}
\label{numerical_simulation_chapter}

\subsection{Finite Element Analysis Model}
\label{FEM_section}

As mentioned in Section \ref{Section_Assumption}, the mechanisms of the formation of pyroshock and ballistic shock are similar.
Both pyroshock and ballistic shock can be produced by the shock induced by metal-to-metal impact.
A frequently-used method to generate pyroshock and ballistic shock in laboratory is called tuned resonant platform technology\cite{Jeong2017}.
For this simple elastic dynamics problem, i.e., simple boundary condition and elastic material behaviour, FEM is a well-established and deterministic method to calculate shock-induced structural response.
Hence, FEM is used to validate the predicted results of shock environment at component interface\cite{Lee2012}. 

\begin{figure}
	\centering
	\includegraphics[width=0.7\textwidth]{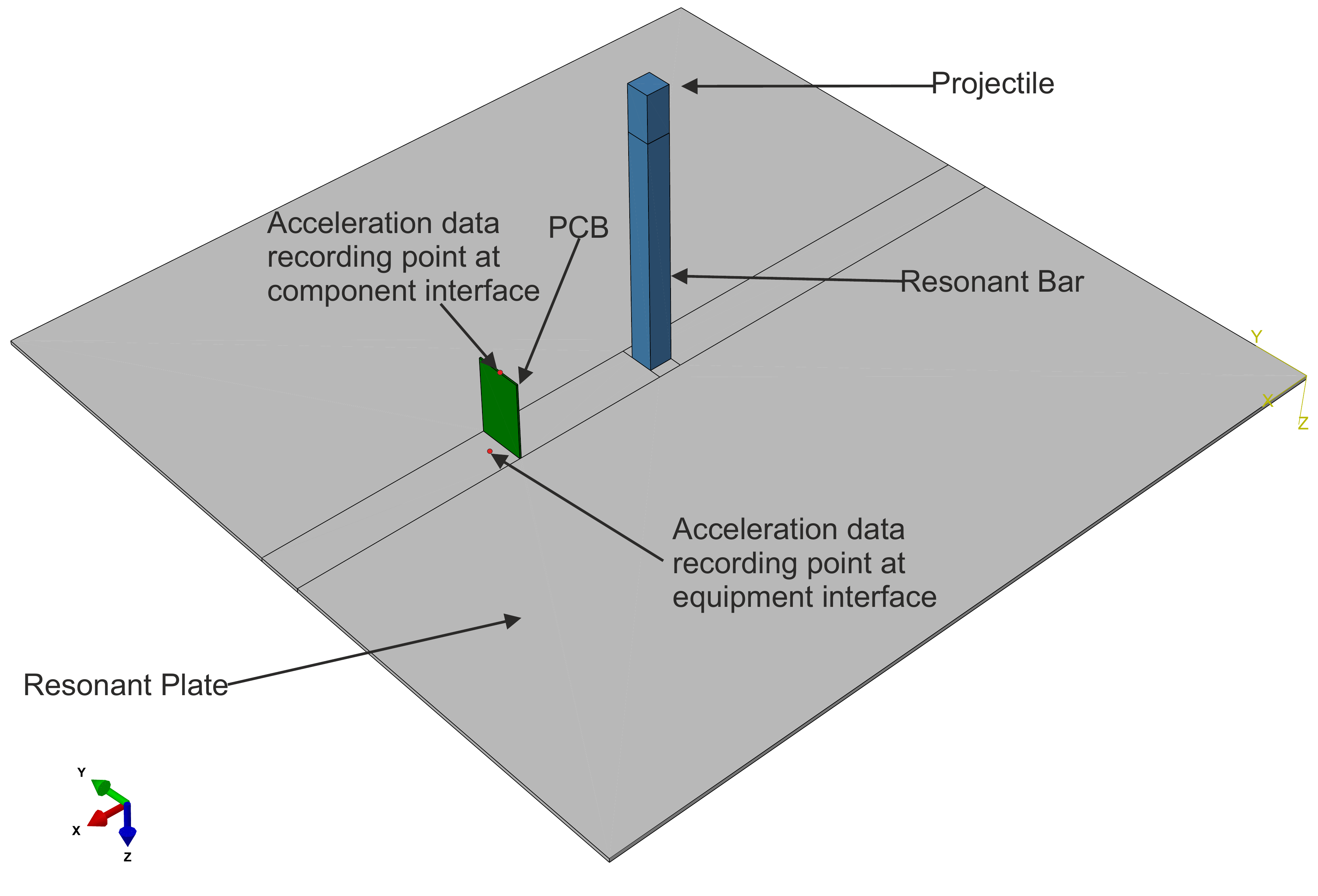}
	\caption{Illustration of tuned resonant platform model}
	\label{FE_Modal}
\end{figure}

The explicit FEM solver in Abaqus 6.14-3 is used. As shown in Fig.\ref{FE_Modal}, a projectile is used to impact the resonant bar to generate a shock environment in a resonant plate.
Two acceleration measuring points are assigned at the top middle (component interface) and the bottom middle (equipment interface) of a plate.
Table \ref{geometric_dimension} shows the geometric dimensions of all parts in FEM model.
To investigate the structural response under shock, a fully-fixed plate representing a simple equipment structure (e.g., printed circuit board (PCB) is used here) is considered for the illustration of the proposed method.
\begin{table}
	\centering
	\begin{threeparttable}
		\caption{Geometric Dimensions of the FEM model}
		\label{geometric_dimension}
		\begin{tabular}{  l  l  l  l  l  }
			\hline
			Part & Resonant Plate & Resonant Bar &  PCB & Projectile\\
			\hline
			Dimension(mm) & $1000\times1000\times5$ & $300\times30\times30$ & $100\times60\times2$ & $60\times30\times30$ \\
			\hline
		\end{tabular}
		%\begin{tablenotes}
		%\item[*] Dimension displays in $mm\times mm\times mm$.
		%\end{tablenotes}
	\end{threeparttable}
\end{table}

General contact algorithm based on ``hard contact" (no penetration) behaviour is used for the impact between the projectile and the resonant bar.
The material data are given in Table \ref{material_property}. The material used in PCB modelling is FR-4, a glass-reinforced epoxy laminate material.
Only elastic model is considered because this analysis only concerns the shock transmissibility in equipment-like structures.
To cover mid-field and far-field shocks, the upper analytic frequency limit in this section is set as 32768 ($2^{15}$) Hz.
\begin{table}
	\centering
	\begin{threeparttable}
		\caption{Material Property}
		\label{material_property}
		\begin{tabular}{l l l l l}
			\hline
			Part & Material & $\rho(kg/m^3)$ & $E(GPa)$ & $\nu$ \\
			\hline
			Resonant Plate & Aluminium & 2800 & 72 & 0.29\\
			Resonant Bar & Steel & 7850 & 200 & 0.25\\
			Projectile & Steel & 7850 & 200 & 0.25\\
			PCB & FR-4 & 1850 & 20 & 0.12 \\
			\hline
		\end{tabular}
		\begin{tablenotes}
			\item Note: $\rho$: density; E: elastic modulus; $\nu$ :Poisson's ratio. 
		\end{tablenotes}
	\end{threeparttable}
\end{table}

To simulate the damping in materials and attenuation in joints, bulk viscosity is introduced\cite{hibbitt2001}.
According to Eqs.(\ref{linear_bulk_viscosity}) and (\ref{quadratic_bulk_viscosity}) given below, linear and quadratic bulk viscosities are introduced to define a bulk viscosity, i.e.
\begin{equation}\label{linear_bulk_viscosity}
p_{bv1}=\rho (b_1 c_d L_e \dot{\epsilon}_{vol})
\end{equation}
\begin{equation}\label{quadratic_bulk_viscosity}
p_{bv2}=\rho (b_2 L_e \dot{\epsilon}_{vol})^2
\end{equation}
where $b_1$ is a non-dimensional linear bulk viscosity parameter, $\rho$ is the current material density, $c_d$ is the current dilatational wave speed, $L_e$ is the element characteristic length, $\dot{\epsilon}_{vol}$ is the volumetric strain-rate and $b_2$ is a non-dimensional quadratic bulk viscosity parameter, $p_{bv1}$ and $p_{bv2}$ are linear and quadratic bulk viscosity pressure, respectively.
The damping ratio from each term can be calculated by
\begin{equation}\label{damping_due_to_bulk_viscosity}
\xi=b_1-b_2^2\frac{L_e}{c_d}min(0,\dot{\epsilon}_{vol})
\end{equation}
In this case, linear bulk viscosity parameter is set to be 0.06 and quadratic bulk viscosity parameter is set to be 1.2.
Based on Eq.(\ref{damping_due_to_bulk_viscosity}), the linear term alone represents 6\% of critical damping, whereas the quadratic term is much smaller and mainly for numerical stability.

\begin{table}
	\centering
	\caption{Determination of element size in FEM}
	\label{element_information}
	\begin{tabular}{l l l l}
		\hline
		Part & Actual Size used & Control Wave & Control Wavelength\\
		\hline
		Resonant Plate &  4 $mm$ & Flexural & 37.5 $mm$ \\
		Resonant Bar &  4 $mm$ & Dilatation & 154.8 $mm$\\
		PCB &  2 $mm$ & Flexural & 19.1 $mm$ \\
		Projectile &  10 $mm$ & Dilatation & 154.0 $mm$\\
		\hline
	\end{tabular}
\end{table}
The element sizes used in this analysis are shown in Table \ref{element_information}.
To minimize the effect of numerical filtering, ECSS suggests that there shall be at least 8 element within the control wavelength considered in the simulation.
The control wave length is determined by the control wave speed and the frequency of the shock, i.e., Eqs.(\ref{wavelength_compression}-\ref{wavelength_flexural}) can be used to determine the control wave length for 1-D dilatation wave, shear wave and flexural wave of frequency $f$, respectively,
\begin{equation}
\lambda=\frac{\sqrt{\frac{E}{\rho}}}{f}
\label{wavelength_compression}
\end{equation}
\begin{equation}
\lambda=\frac{\sqrt{\frac{G}{\rho}}}{f}
\label{wavelength_shear}
\end{equation}
\begin{flalign}
\text{and} && \lambda=(\frac{2\pi}{f})^{\frac{1}{2}}(\frac{Et^2}{12\rho})^{\frac{1}{4}} &&
\label{wavelength_flexural}
\end{flalign}
where $G$ is the shear modulus of the material; t is the thickness of the 2D structure; upper limit of the shock frequency $f=32768$ Hz is used to calculate control wavelengths in Table \ref{element_information}. The actual element size should be less than $\frac{1}{8}$ of the control wave length.

In practice, the actually-used mesh sizes should be geometrically compatible, i.e., the element sizes for resonant bar and projectile cannot be too different from the actual mesh size for other parts.
Otherwise over distorted mesh grids may be generated.
The actual element mesh sizes shown in Table \ref{element_information} for different parts of the FEM model were selected to meet both ECSS requirement and the geometrical compatibility.

The time step for FEM simulation needs to be small enough to satisfy the sampling requirement of all analysed frequency range.
From the FEM explicit integration scheme, the numerical stability is ensured by the verification of Courant condition, i.e.
\begin{equation}
\triangle t_{max}\leq \frac{\triangle x}{c}
\end{equation}
where $\triangle x$ is the minimum element size; $c$ is the speed of control wave; and $\triangle t_{max}$ is the maximum time step in simulation.

This condition imposes that the maximum time step $\triangle t_{max}$ is smaller than the travelling time duration of the fastest stress wave (dilatation wave) over the minimum element size $\triangle x$ inside the whole FEM model.
In this section, time step of $0.29\,\mu s$ is adopted.

\begin{figure}
	\centering
	\begin{subfigure}[b]{0.5\textwidth}
		\includegraphics[width=\textwidth]{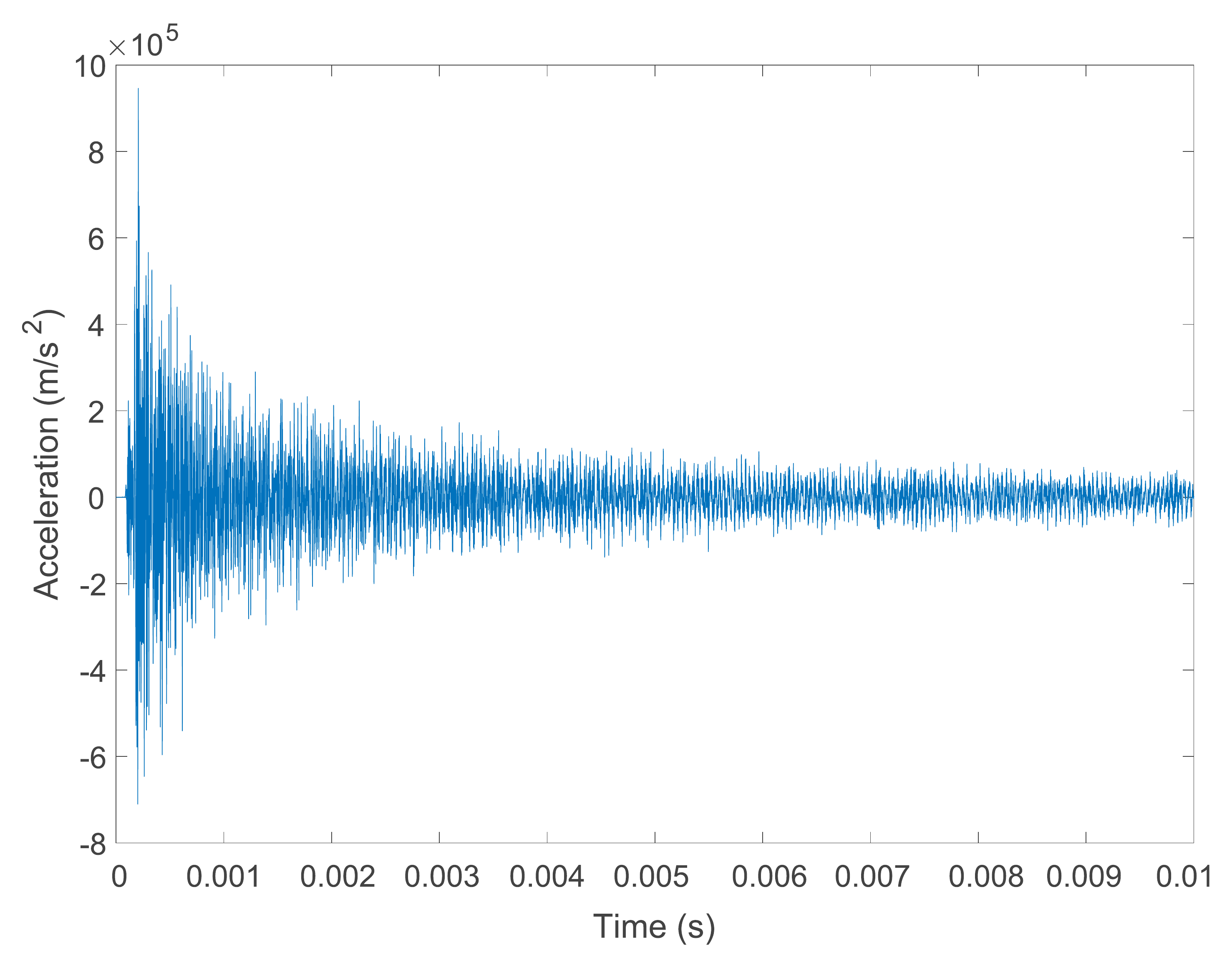}
		\caption{}
		\label{time_history_equipment_IF}
	\end{subfigure}
	\begin{subfigure}[b]{0.5\textwidth}
		\includegraphics[width=\textwidth]{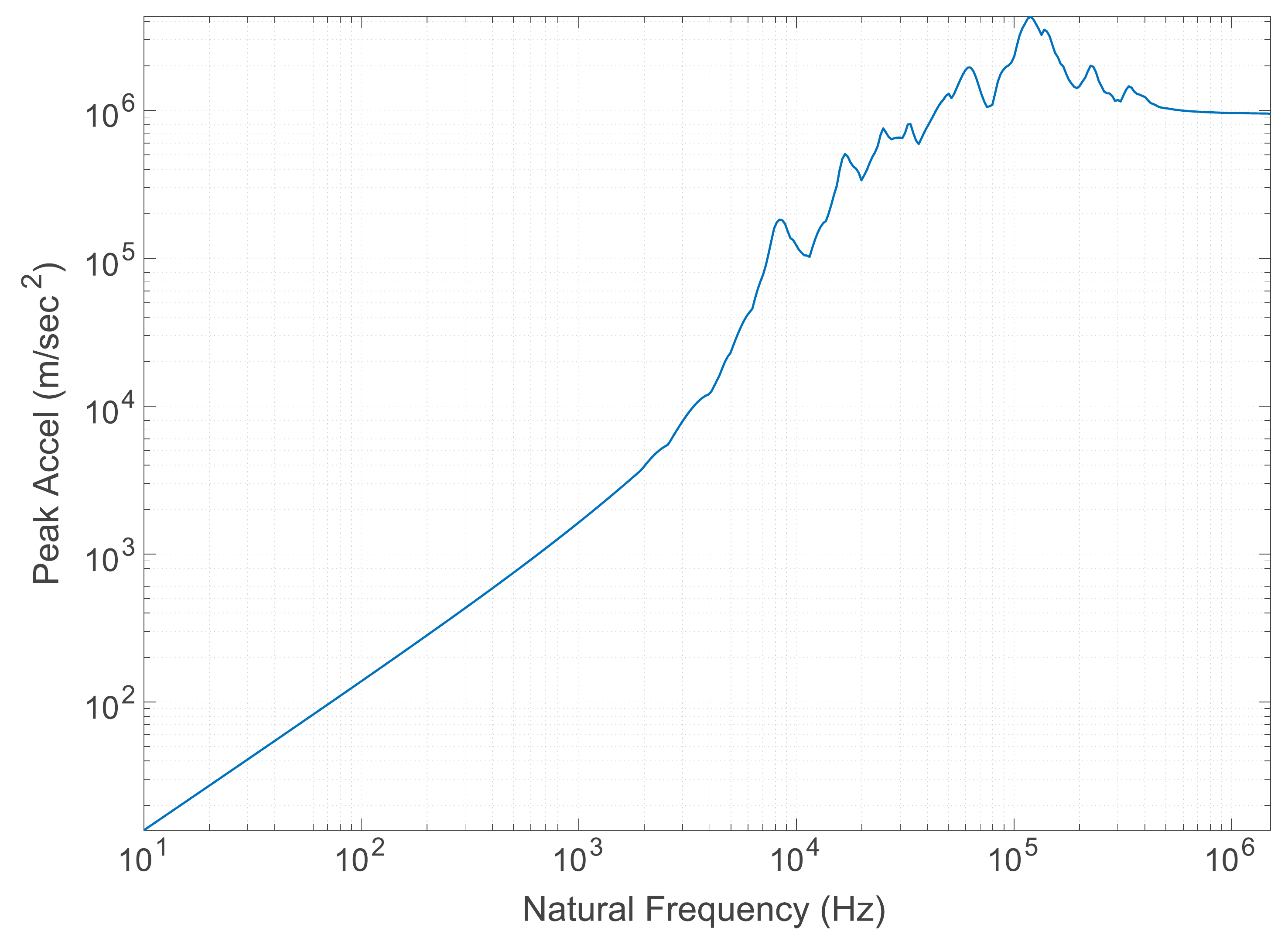}
		\caption{}
	\end{subfigure}
	\caption{(a) Time history, and (b) SRS, of the acceleration signal at equipment interface in Fig.\ref{FE_Modal} at impact speed of 20 m/s}
	\label{FEM_shock_environment}
\end{figure}

As shown in Fig.\ref{FE_Modal}, acceleration-time history signals are collected at both equipment and component interfaces.
The acceleration signal at equipment interface represents the shock environment, whose time history and SRS are displayed in Fig.\ref{FEM_shock_environment} for projectile impact speed of 20 m/s in Fig.\ref{FE_Modal}.

\subsection{Analysis with LPSRS method}
\begin{figure}
	\centering
	\includegraphics[width=0.6\textwidth]{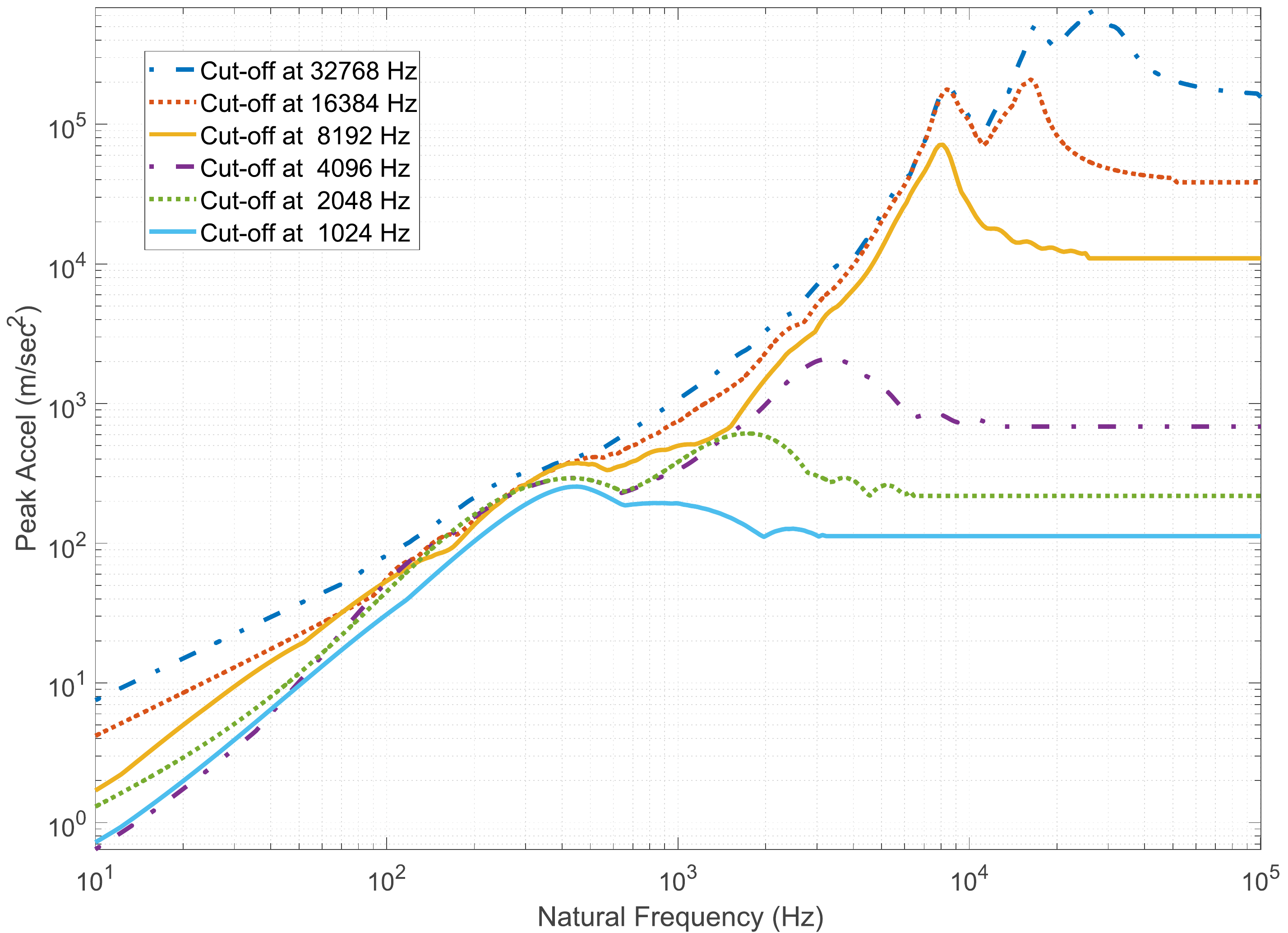}
	\caption{The LPSRS of the acceleration signal at equipment interface in Fig.\ref{FEM_shock_environment}(a)}
	\label{LPSRS of acceleration signal at equipment interface}
\end{figure}

\begin{figure}
	\centering
	\includegraphics[width=0.8\textwidth]{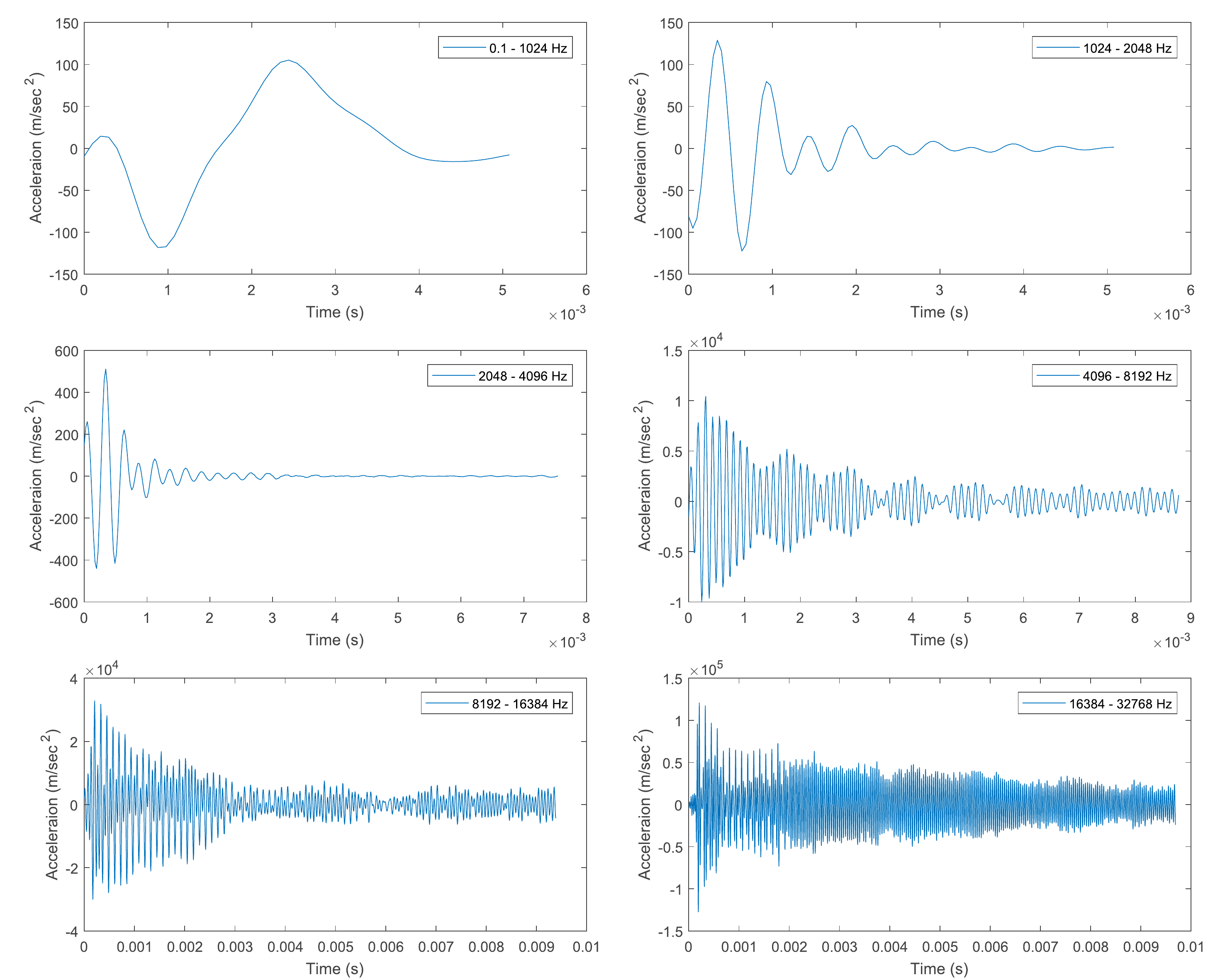}
	\caption{Acceleration-time history at an equipment interface filtered by FIR band-pass filter designed with Hamming window(band range is shown in each figure) for the shock shown in Fig.\ref{FEM_shock_environment}(a).}
	\label{bandpass_signals}
\end{figure}

A LPSRS plot of the shock acceleration-time history in Fig.\ref{FEM_shock_environment}(a) is displayed in Fig.\ref{LPSRS of acceleration signal at equipment interface}.
The frequency cut-off in LPSRS analysis is conducted from 1024 Hz to 32768 Hz every each octave frequency.
Filter analysis for frequency lower than 1024 Hz is unnecessary, because the amplitude of shock signal filtered by a 1024 Hz low-pass filter is almost zero.
It is found that the amplitude difference between two neighbouring platforms of LPSRS curve approaches to the temporal maximum acceleration amplitude of corresponding band-pass signal, which supports the linear superposition assumption (Eq.(\ref{linear_superpostion_assumption})).
This conclusion can be simply validated by comparing amplitude read from LPSRS and the corresponding band-pass-filtered signal.
Here a band-pass FIR filter designed with Hamming window was used to extract amplitude of filtered signal $A_i$, as illustrated in Fig.\ref{bandpass_signals}.
The magnitude and phase response of band-pass filter used are not shown here, as they depend on filtering frequency range.
But they have similar properties of the low-pass filter shown in Fig.\ref{bode_plot}.
Phases of filtered signals usually delay several periods.
A compensation is necessary for the delay introduced by such FIR filter in order to locate the peak time more accurately, which is achieved by a function embedded in Matlab to shift the signal in time domain.
Table \ref{comparison of amplitudes} compares the platform amplitude difference obtained from LPSRS and the maximum amplitude of band-pass-filtered signal.
Amplitudes difference read directly from the neighbouring LPSRS platforms are close to the maximum amplitudes obtained from band-pass-filtered signals.
The amplitude ratio is only 0.13 dB in frequency band between 4096 - 8192 Hz, and the largest amplitude ratio is 1.72 dB, which adequately demonstrate the validity of the linear superposition assumption in pyroshock and ballistic shock applications.

\begin{table}
	\centering
	\small
	\caption{Comparison of platform amplitudes difference read from LPSRS and maximum amplitude from band-pass-filtered signal}
	\label{comparison of amplitudes}
	\begin{tabular}{r r r r}
		\hline
		Frequency Band & $A_i (m/s^2)$ & $A^i-A^{i-1} (m/s^2)$ & Amplitude Ratio\\
		\hline
		0 - 1024 Hz& 118.1 & 112.5 & 0.42 dB\\
		1024 - 2048 Hz& 128.9 & 105.8 & 1.72 dB\\
		2048 - 4096 Hz& 509.4 & 466.7 & 0.76 dB\\
		4096 - 8192 Hz& 10420 & 10265 & 0.13 dB\\
		8192 - 16384 Hz& 32860 & 27440 & 1.57 dB\\
		16384 - 32768 Hz& 127400 & 117810 & 0.68 dB\\
		\hline
	\end{tabular}
\end{table}

\begin{figure}
	\centering
	\includegraphics[width=0.25\textwidth]{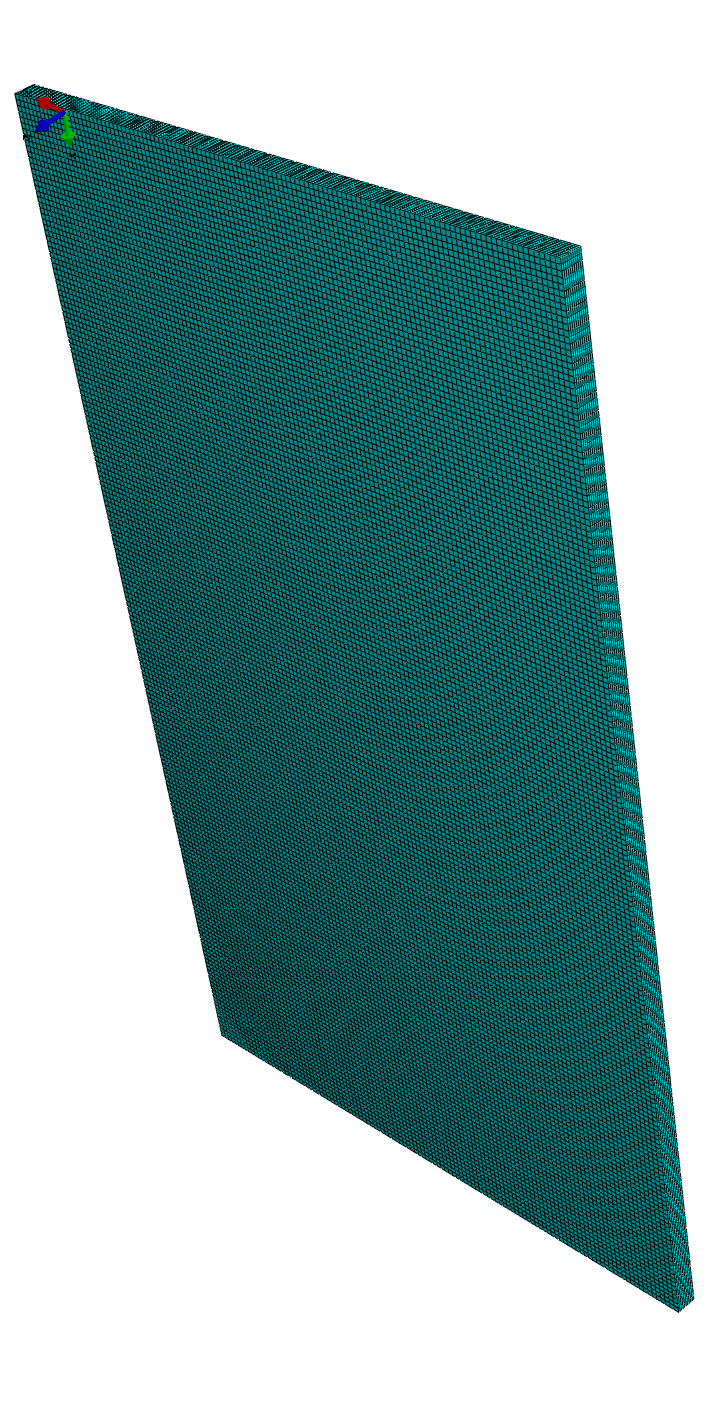}
	\caption{FEM modal used for frequency extraction}
	\label{frequency_extraction_FEM_modal}
\end{figure}

In this section, out-of-plane ($x$-axis in Fig.\ref{FE_Modal}) shock response of response plate (PCB) was analysed.
To evaluate the shock environment at component interface with Eq.(\ref{LPSRS_algorithm}), some basic modal information of the response plate is required.
Lanczos frequency solver based on linear perturbation procedure is used to extract natural frequencies $f_n$, modal participation factors $P_n$ and mode shape $\phi_n$ from PCB.
All modes within 32768 Hz and with effective mass higher than $5\times 10^{-5}\,g$ are listed in Table \ref{modal_information}.
The sum of extracted effective masses is 21.3 g, which is already 96\% of the total mass (22.2 g) of PCB\cite{irvine2013}.
Therefore, it is sufficient to describe the PCB with such modal information. To meet mesh requirement, mesh size is reduced gradually until the natural frequencies is converged.
In this case, mesh size used is 0.2 mm in Fig.\ref{frequency_extraction_FEM_modal}, where 267531 nodes and 240000 elements in total are created.

\begin{table}
	\centering
	\caption{Modal Information For PCB}
	\label{modal_information}
	\begin{tabular}{l l l l l}
		\hline
		Order & $f_n$(Hz) & $P_n$ & $\phi_n$ & Effective Mass (kg)\\
		\hline
		1     & 103   & -0.1170  & -13.44 & $1.36\times10^{-2}$  \\
		3     & 646   & 0.0647  & -13.58 & $4.19\times10^{-3}$  \\
		5     & 1800  & 0.0375  & 15.54  & $1.40\times10^{-3}$  \\
		10    & 3528  & 0.0271  & -12.52 & $9.33\times10^{-4}$  \\
		14    & 5805  & -0.0211 & -12.87 & $4.45\times10^{-4}$  \\
		20    & 8621  & 0.0172  & -13.03 & $2.98\times10^{-4}$  \\
		28    & 11958 & -0.0144 & -9.88  & $2.06\times10^{-4}$  \\
		35    & 15805 & 0.0127  & -12.77 & $1.61\times10^{-4}$  \\
		46    & 20128 & -0.0112 & -14.08 & $1.26\times10^{-4}$  \\
		57    & 24907 & -0.0100 & 13.11  & $1.01\times10^{-4}$  \\
		69    & 30123 & 0.0091 & 11.07  & $8.27\times10^{-5}$ \\
		\hline
	\end{tabular}
\end{table}

\begin{figure}
	\centering
	\includegraphics[width=0.6\linewidth]{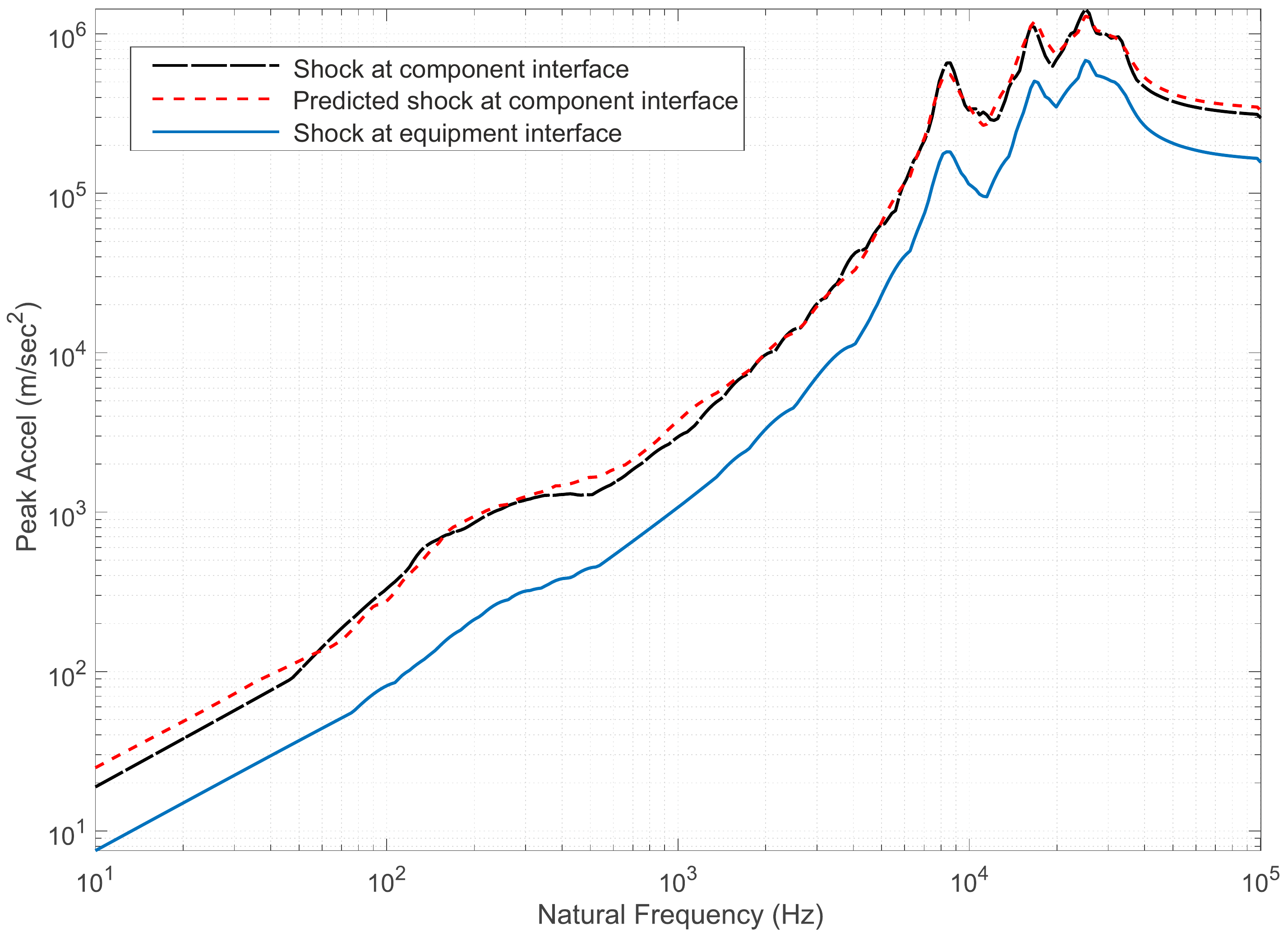}
	\caption{Comparison of predicted shock by LPSRS method and numerically measured shock by FEM}
	\label{prediction_comparison}
\end{figure}

Based on Eq.(\ref{LPSRS_algorithm}), an algorithm is implemented in Matlab, which, together with basic modal information in Table \ref{modal_information}, LPSRS and SRS curves at any component interface can be obtained from LPSRS at a given equipment interface.
Fig.\ref{prediction_comparison} shows the predicted absolute acceleration SRS at component interface using LPSRS at equipment interface in Fig.\ref{LPSRS of acceleration signal at equipment interface} and modal information in Table \ref{modal_information}, together with the corresponding SRS calculated from the numerically measured acceleration-time history data at component interface using ABAQUS.
It shows that the SRS predicted from LPSRS method agrees well with the SRS calculated from FEM.
It can be seen that the shock severity at component interface is much higher than the shock severity at equipment interface.
After it is transmitted from equipment interface to component interface, the SRS at component interface could be more than five times severer than that at equipment interface.
If these differences are not taken into consideration, components may be overloaded and fail during a mission because of the under-estimation of shock severity at sensitive component interface.

\begin{figure}
	\centering
	\begin{subfigure}[b]{0.5\linewidth}
		\includegraphics[width=\textwidth]{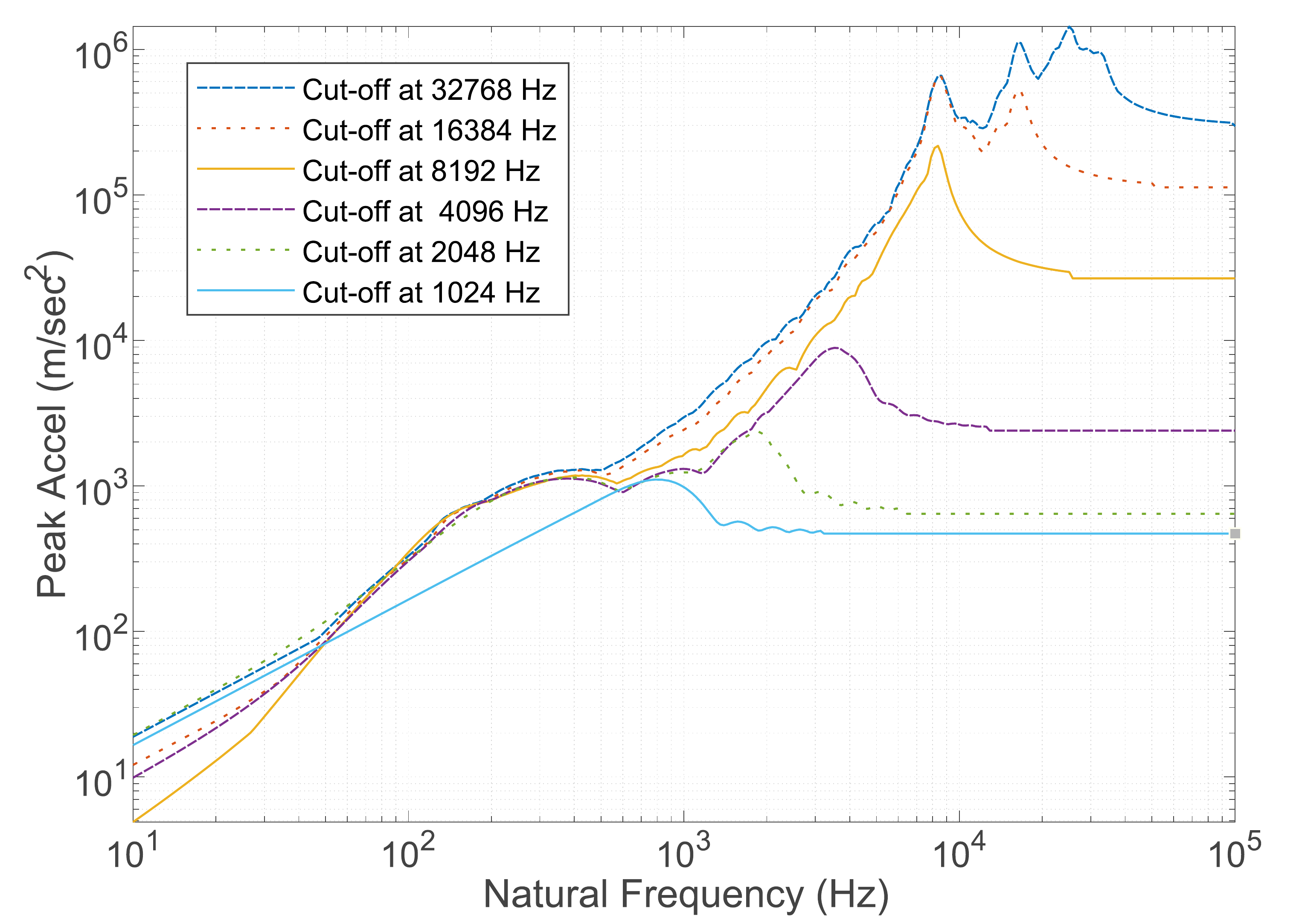}
		\caption{}
	\end{subfigure}
	\begin{subfigure}[b]{0.5\textwidth}
		\includegraphics[width=\linewidth]{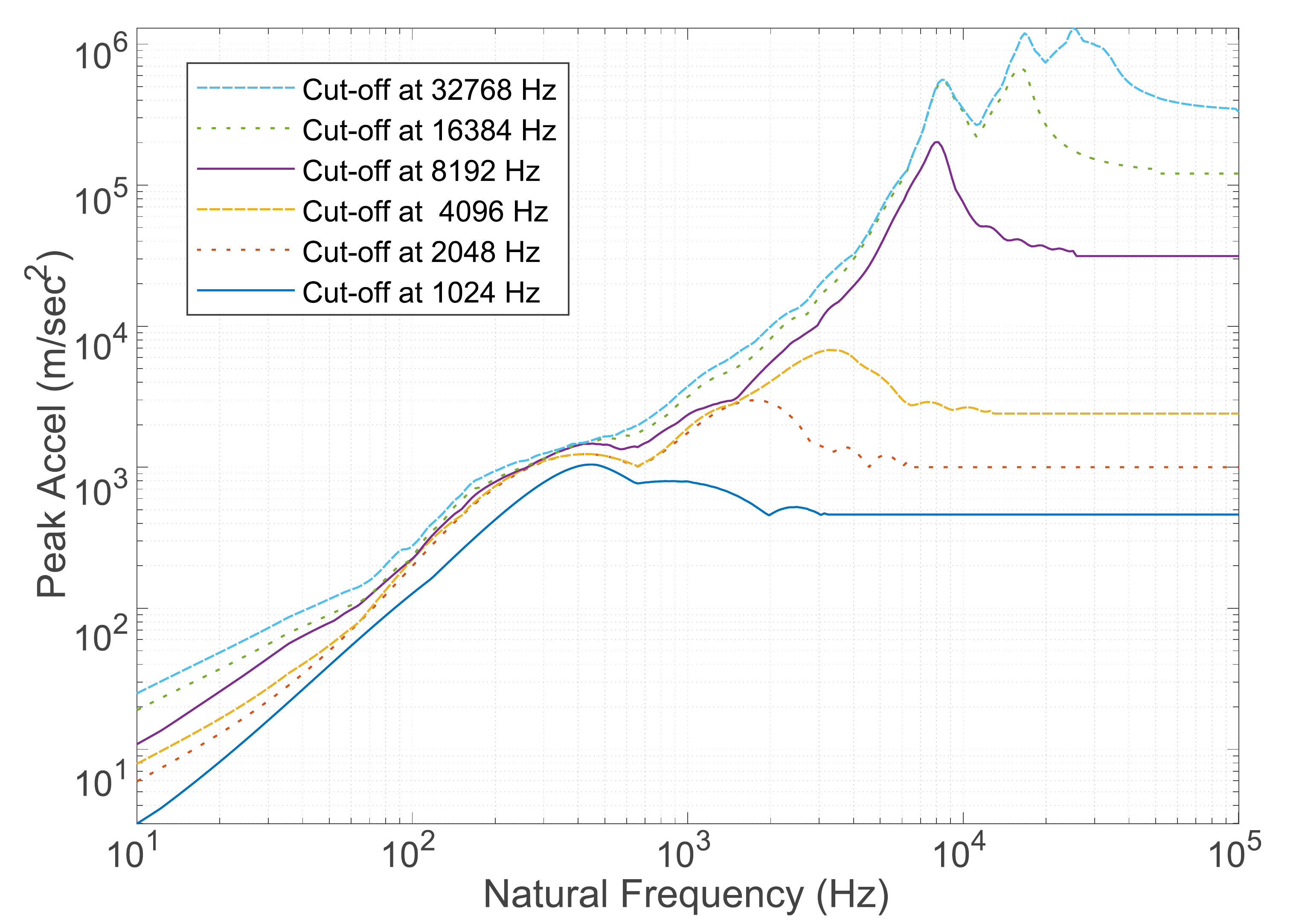}
		\caption{}
	\end{subfigure}
	\caption{Comparison of (a) FEM-measured LPSRS, and (b) Predicted LPSRS, at component interface for the shock excitation at equipment interface in Fig.\ref{FEM_shock_environment}(a).}
	\label{LPSRS_comparison}
\end{figure}

Another advantage of LPSRS method is that the prediction method does not lose any information during the process of shock transmission.
With Eq.(\ref{frequency_information_hold}), not only total response $a_f$, but also the amplitude reflecting frequency information, $a_{if}$, can be calculated for the shock transmitted to a component interface.
Fig.\ref{LPSRS_comparison} compares the predicted LPSRS and the LPSRS measured directly from the shock signal at a component interface using FEM.
Platform amplitudes read directly from the predicted and collected LPSRS curves are given in Table \ref{comparison_amplitude_LPSRS}, where differences are acceptable.
This implies that LPSRS method has the potential to calculate the shock transmissibility inside a multilevel structure, e.g. spacecraft, although further study is still need to estimate its accuracy.
\begin{table}
	\centering
	\begin{threeparttable}
		\caption{Comparison of platform amplitude differences read from predicted LPSRS and FEM-measured LPSRS}
		\label{comparison_amplitude_LPSRS}
		\begin{tabular}{r r r r}
			\hline
			Cut-off Frequency & ${a^i_{predicted}}^a\:(m/s^2)$ & ${a^i_{collected}}^b\:(m/s^2)$ & Amplitude Ratio \\
			\hline
			1024 Hz& 462 & 470 & 0.15 dB  \\
			2048 Hz& 1001 & 643 & 3.84 dB \\
			4096 Hz& 2401 & 2398 & 0.01 dB  \\
			8192 Hz& 31420 & 26660 & 1.43 dB \\
			16384 Hz& 120800 & 112600 & 0.61 dB \\
			32768 Hz& 330900 & 296800 & 0.94 dB\\
			\hline
		\end{tabular}
		\begin{tablenotes}
			\item[] {\footnotesize $^a$ Platform amplitude read form predicted LPSRS; $^b$ Platform amplitude read form FEM-measured LPSRS}
		\end{tablenotes}
	\end{threeparttable}
\end{table}

\subsection{Comparison with ECSS's transmissibility evaluation method}
According to ECSS, methods for predicting shock severity level at sensitive component interface have not been fully developed.
It mainly relies on rule-of-thumb(\cite{ECSS2015}, p.463).
The shock transmissibility function $TF^{shock}(f)$ between equipment and component interfaces is used to estimate the SRS curve at component interface.
Normally $TF^{shock}(f)$ is obtained from available sweep sine survey test, with following steps:
\begin{itemize}
	\item[(i)] Extracting the transmissibility frequency response function, i.e. $FRF^{sine}(f)$, between equipment and component interfaces from a sweep sine survey test, where equipment should be mounted on a rigid interface under a uniformly distributed sine base excitation with varied frequency.
	\item[(ii)] The bounds of $TF^{shock}(f)$ between equipment and component interfaces below 2000 Hz can be obtained by:
	\begin{equation}
	\sqrt{FRF^{sine}(f)}\leq TF^{shock}(f)\leq\sqrt{2\times FRF^{sine}(f)}
	\end{equation}
	If the $FRF^{sine}(f)$ is the result of a FEM simulation, the bounds of $TF^{shock}(f)$ are:
	\begin{equation}
	\sqrt{FRF^{sine}(f)}\leq TF^{shock}(f)\leq 2\sqrt{FRF^{sine}(f)}
	\end{equation}
	\item[(iii)] In frequency band defined between 2000 Hz and the transition frequency, the transmissibility between equipment and component interfaces is assumed to be 6 dB, so that $TF^{shock}(f)\approx2$.
	\item[(iv)] Beyond the transition frequency and up to the maximum frequency of SRS at equipment interface, a decreasing corridor is proposed, which was not specified in \cite{ECSS2015}.
	\item[(v)] SRS at component interface can be obtain by
	\begin{equation}
	SRS \text{ at component interface = }SRS \text{ at equipment interface}\times TF^{shock}(f)
	\end{equation}
\end{itemize}

\begin{figure}
	\centering
	\includegraphics[height=0.9\textheight]{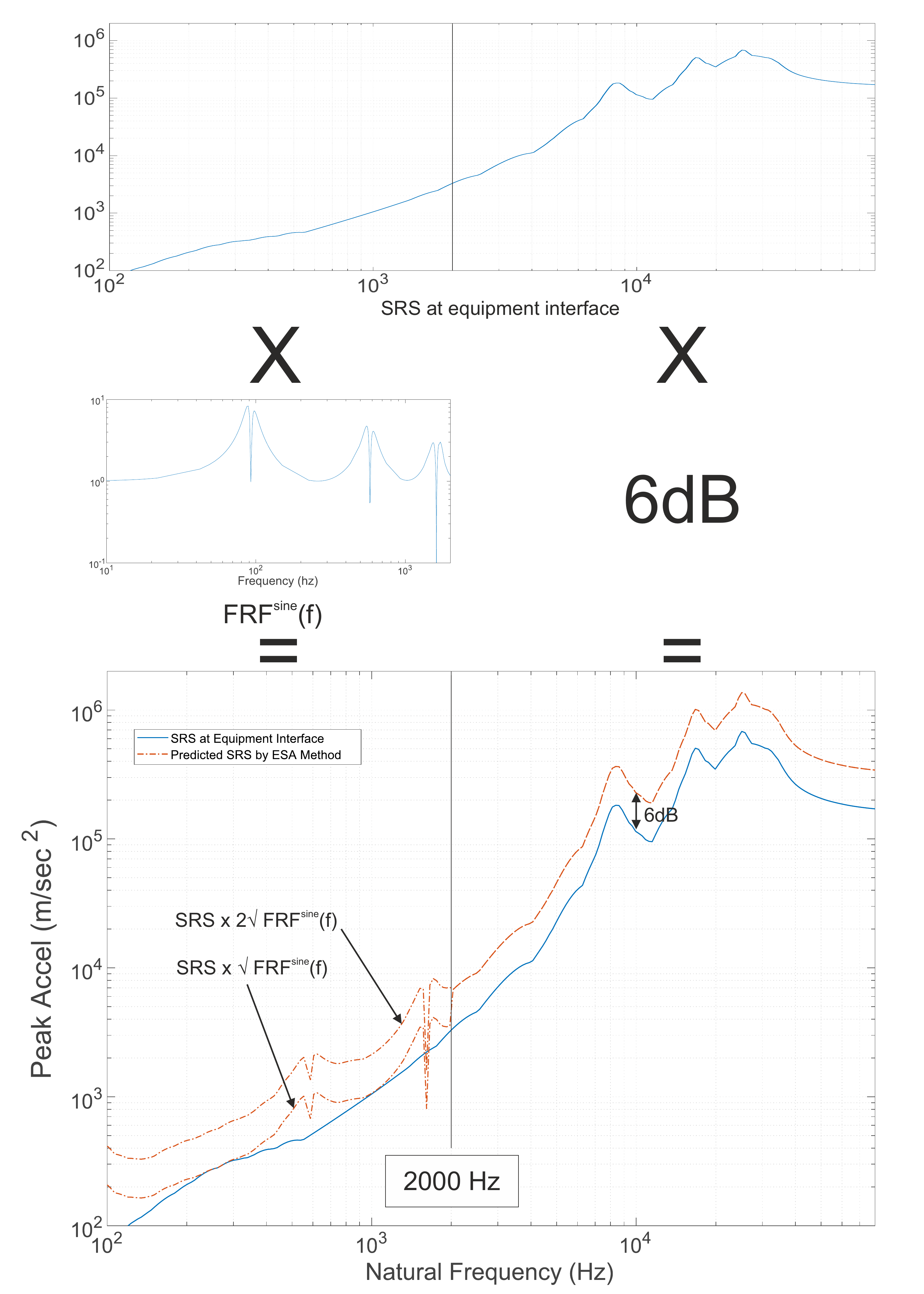}
	\caption{Shock transmissibility prediction method by ECSS (when $FRF^{sine}$ is based on FEM simulation)}
	\label{flowchart_ESA}
\end{figure}

In this example, $FRF^{sine}(f)$ between equipment and component interfaces is extracted from the same FEM model as shown in Fig.\ref{frequency_extraction_FEM_modal} with ``Steady-state dynamics, Modal" procedure.
The model is under a series of sine excitation on a rigid interface.
As an illustration of the prediction procedure, a diagram is depicted in Fig.\ref{flowchart_ESA}.
The $TF^{shock}(f)$ below 2000 Hz is between the square root of the sine transmissibility and the twice the square root of sine transmissibility (when the $FRF^{sine}(f)$ is the result from FEM simulation).
After 2000 Hz, a 6 dB corridor is applied until transition frequency.
In this case, transition frequency (i.e. around 30 kHz) is close to the maximum cut-off frequency in LPSRS, so no decreasing corridor is applied.

\begin{figure}
	\centering
	\includegraphics[width=0.6\linewidth]{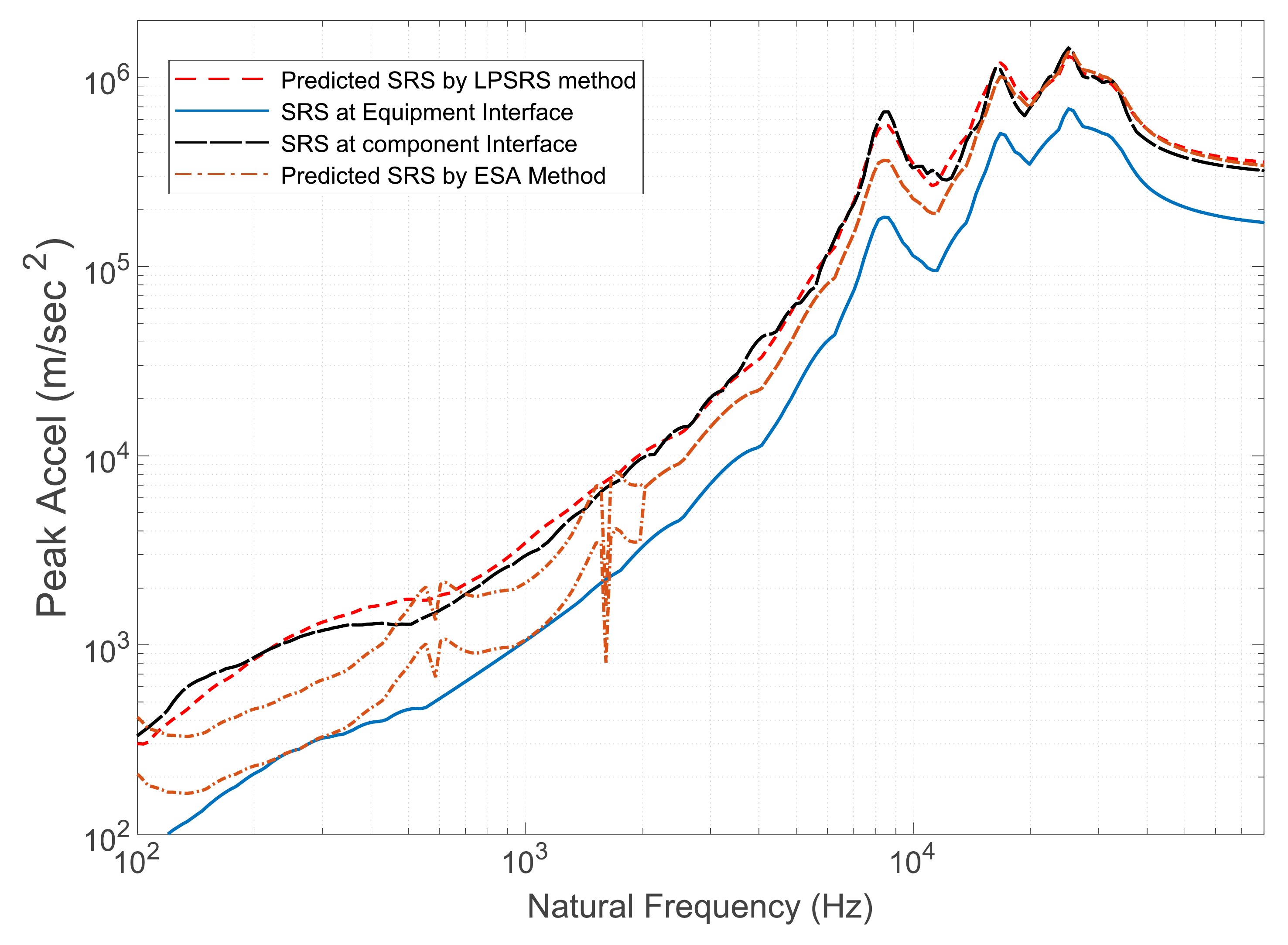}
	\caption{Comparison of transmissibility by LPSRS and ECSS method}
	\label{result_comparison}
\end{figure}

A comparison between SRS curves predicted by ECSS method and the SRS predicted by LPSRS method is shown in Fig.\ref{result_comparison}.
In general, the SRS predicted by LPSRS method at component interface is better than the SRS predicted by ECSS method.
Although the ECSS method provides a range between upper and lower bounds below 2000 Hz, it still underestimates the shock severity at component interface even with its upper bound envelope.
The SRS at component interface is sometimes twice higher than the upper envelope of the predicted SRS by ECSS.
From 2000 Hz to 15 kHz, shock severity is still underestimated with the proposed 6 dB corridor by ECSS, while SRS predicted by LPSRS method is almost the same as the SRS measured (numerically) at the component interface.
After 15 kHz, both the 6 dB corridor and the LPSRS method can meet the measured SRS at component interface.

\section{Conclusions}
According to pyroshock and ballistic shock generation mechanism, this study considers pyroshock and ballistic shock as the superposition of damped harmonic wave with wide range of frequencies.
According to the linear superposition assumption, a shock analysis method based on low-pass filter and shock response spectrum method (LPSRS) is proposed.
LPSRS can show the contribution of the frequency distribution of the shock environment at an input location of an equipment structure.
A shock transmissibility calculation is proposed based on LPSRS and basic modal information of the equipment structure to predict shock environment (LPSRS or SRS) at any given position on the structure. The predictive method is validated by FEM simulation.

Another significance of LPSRS method is its benchmark function to characterize shock severity.
Traditional SRS can only calculate upper bound structural response under shock excitation.
Different shocks may have the same SRS curve, and thus, their shock severities are considered to be the same, and hence, ``equivalence" between shocks can be established.
However, when referring to the response of a substructure, i.e., a sensitive component, the response of the substructure may be different for shocks with the same SRS.
Therefore, the general effectiveness of SRS method as a measure of shock severity is questionable. 

While calculating the shock response of a structure, LPSRS method can maintain its frequency distribution and contribution.
As long as LPSRS at equipment interface is available, shock environment at a substructure interface can be determined by SRS or LPSRS, with which it is possible to calculate the response of the substructure.
Further more, LPSRS curves contain more essential shock characters than SRS curve, and therefore LPSRS is a better choice for the representation of a shock environment. 

\section*{References}
%\begin{thebibliography}{00}
%% \bibitem{label}
%% Text of bibliographic item
\bibliography{paper}
%\end{thebibliography}

\setcounter{figure}{0}
\appendix
\section{Response of a SDOF oscillator under a rectangular pulse}\label{section_appendix}
Here a rectangular pulse force is used as an example of pulse loading, i.e.
\begin{figure}[hbtp]
	\centering
	\includegraphics[scale=1]{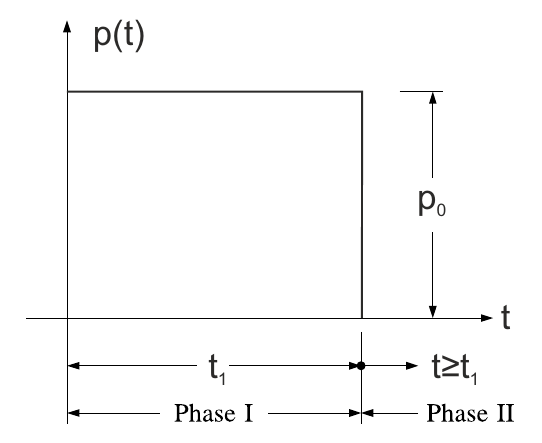}
	\caption{Rectangular pulse}
	\label{rectangular_impluse}
\end{figure}

\begin{equation}
p(t) =
\begin{cases}
0 & t<0\\
p_0  & 0 \leq t \leq t_1\\
0 & t>t_1
\end{cases}
\end{equation}
where $t_1$ is the loading duration; $p_0$ is the amplitude of the loading.

The displacement $u(t)$ and acceleration responses $\ddot{u}(t)$ of one of the response modes of a structure under rectangular pulse can be represented by the response of a SDOF model, which can be divided into two phases\cite{Clough2003}, i.e. the forced response phase during the loading period
\begin{equation}
u(t)=\frac{p_0}{k}\{1-e^{-\xi \omega t}[\cos(\omega_Dt)+\frac{\xi}{\sqrt{1-\xi^2}}\sin(\omega_Dt)]\}, \quad 0\leq t\leq t_1
\end{equation}

\begin{multline}
\label{forced}
\ddot{u}(t)=\frac{1}{k\sqrt{1-\xi^2}} e^{-t \xi \omega }p_0 (\xi^2\sqrt {1 - \xi^2}\;\omega^2\cos (t \omega_D)-2\xi^2 \omega \omega_D \cos(t \omega_D)-\sqrt{1-\xi^2}\;\omega_D^2 \cos(t \omega_D)\\
+\xi^3 \omega^2 \sin(t \omega_D)+2 \xi \sqrt{1-\xi^2}\;\omega \omega_D \sin(t \omega_D)-\xi \omega_D^2 \sin(t \omega_D)), \quad 0\leq t\leq t_1
\end{multline}
and the subsequent free vibration phase can be obtained by
\begin{multline}
u(t)=\frac{1}{k\omega_D \sqrt{1-\xi^2}}e^{-t \xi \omega}p_0(-\sqrt{1-\xi^2}\;\omega_D \cos(t\omega_D)+e^{t_1 \xi \omega}\sqrt{1-\xi^2}\;\omega_D \cos((t-t_1)\omega_D)\\
-\xi \omega_D \sin(t\omega_D)+e^{t_1 \xi \omega}\xi \sqrt{1-\xi^2}\;\omega \sin((t-t_1)\omega_D)), \quad t>t_1 
\end{multline}
\begin{multline}
\label{free}
\ddot{u}(t)=\frac1{k\sqrt{1-\xi^2}\;\omega_D}e^{-t\xi\omega}p_0(-\xi^2\sqrt{1-\xi^2}\;\omega^2\omega_D \cos(t\omega_D)+2\xi^2\omega\omega_D^2\cos (t\omega_D)+\sqrt{1-\xi^2}\;\omega_D^3\cos( t\omega_D)\\
-e^{t_1\xi\omega}\xi^2\sqrt{1-\xi^2}\;\omega^2\omega_D({\cos}(t-t_1)\omega_D)-e^{t_1\xi\omega}\sqrt{1-\xi^2}\;\omega_D^3\cos((t-t_1)\omega_D)-\xi^3\omega^2\omega_D\sin(t\omega_D)\\
-2\xi\sqrt{1-\xi^2}\;\omega\omega_D^2 \sin(t\omega_D)+\xi\omega_D^3 \sin(t\omega_D)+e^{t_1\xi\omega}\xi^3\sqrt{1-\xi^2}\;\omega^3 \sin((t-t_1)\omega_D)+\\e^{t_1\xi\omega}\xi\sqrt{1-\xi^2}\;\omega\omega_D^2 \sin((t-t_1))\omega_D), \quad t>t_1
\end{multline}
where $k$ is the stiffness of the spring of SDOF oscillator; $\omega$ is oscillator's circular natural frequency; damped natural frequency $\omega_D=\sqrt{1-\xi^2}\;\omega$ where $\xi$ is the damping ratio of the SDOF oscillator.

In low damping ratio condition, i.e. $\xi\leq0.05$, following approximation can be adopted
\begin{equation}
\omega_D\approx\omega
\end{equation}
\begin{equation}
\xi^2\approx 0
\end{equation}
and when these approximations are applied, Eq.(\ref{forced}) and Eq.(\ref{free}) can be simplified to
\begin{equation}
\ddot{u}(t)=\frac{p_0 \omega^2 e^{-\xi \omega t} \cos (\omega t)}{k},\quad 0\leq t\leq t_1
\end{equation}
and
\begin{equation}
\ddot{u}(t)=\frac{p_0 \omega^2 \left(e^{-\xi \omega t} \cos (t \omega)-e^{-\xi \omega (t-\text{t1})} \cos (\omega (t-\text{t1}))\right)}{k},\quad t>t_1
\end{equation}
respectively.

\end{document}